\documentclass[prl,twocolumn,english,superscriptaddress,longbibliography]{revtex4-1}

\usepackage[T1]{fontenc}
\usepackage[latin9]{inputenc}
\usepackage{color}
\usepackage{latexsym}
\usepackage{float}
\usepackage{amsmath}
\usepackage{amsfonts}
\usepackage{graphicx}
\usepackage{times}   %% Times Roman font
\usepackage{esint}
\usepackage{subfigure}
\usepackage{verbatim}
\usepackage{algorithm}
\usepackage{algorithmicx}
\usepackage{algpseudocode}
\usepackage{soul}
\usepackage{xcolor,colortbl}
\usepackage[unicode=true,pdfusetitle,
 bookmarks=false,colorlinks=true,citecolor=blue,urlcolor=blue,linkcolor=red]{hyperref}

\makeatletter

\@ifundefined{textcolor}{}
{%
 \definecolor{BLACK}{gray}{0}
 \definecolor{WHITE}{gray}{1}
 \definecolor{RED}{rgb}{1,0,0}
 \definecolor{GREEN}{rgb}{0,1,0}
 \definecolor{BLUE}{rgb}{0,0,1}
 \definecolor{CYAN}{cmyk}{1,0,0,0}
 \definecolor{MAGENTA}{cmyk}{0,1,0,0}
 \definecolor{YELLOW}{cmyk}{0,0,1,0}
\definecolor{TABLE}{rgb}{0,1,0}
}

\@ifundefined{date}{}{\date{}}
\AtBeginDocument{
  
}
\makeatother

\newcommand{\SAVE}[1]{}

\newcommand{\mc}[2]{\multicolumn{#1}{|c|}{#2}}

\begin{document}
\renewcommand\abstractname{}

\title{Modern Approaches to Exact Diagonalization and Selected Configuration Interaction with the Adaptive Sampling CI Method}
\author{Norm M. Tubman, C. Daniel Freeman, Daniel S. Levine, Diptarka Hait, Martin Head-Gordon, K. Birgitta Whaley}
\affiliation{Kenneth S. Pitzer Center for Theoretical Chemistry, Department of Chemistry, University of California, Berkeley, California 94720, USA and Chemical Sciences Division, Lawrence Berkeley National Laboratory Berkeley, California 94720, USA}
\date{\today}
\begin{abstract} 
Recent advances in selected CI, including the adaptive sampling configuration interaction (ASCI) algorithm and its heat bath extension,
have made the ASCI approach competitive with the most accurate techniques available, and hence an increasingly powerful tool in solving quantum Hamiltonians.  In this work, we show that a useful paradigm for generating efficient selected CI/exact diagonalization algorithms is driven by fast sorting algorithms, much in the same way iterative diagonalization is based on the paradigm of matrix vector multiplication.  We present several new algorithms for all parts of performing a selected CI, which includes new ASCI search, dynamic bit masking, fast orbital rotations, fast diagonal matrix elements, and residue arrays.   The algorithms presented here are fast and scalable, and  we find that because they are built on fast sorting algorithms they are more efficient than all other approaches we considered.
After introducing these techniques we present ASCI results applied to a large range of systems and basis sets in order to demonstrate the types of simulations that can be practically treated at the full-CI level with modern methods and hardware, presenting double- and triple-zeta benchmark data for the G1 dataset.  The largest of these calculations is Si$_{2}$H$_{6}$ which is a simulation of 34 electrons in 152 orbitals.  We also present some preliminary results for fast deterministic perturbation theory simulations that use  hash functions to maintain high efficiency for treating large basis sets. 

\end{abstract}
\maketitle

\newpage

\section{Introduction}
\label{sec:intro}
%%%%%%%%%%%%%%%%%%%%%%
Selected configuration interaction techniques (SCI) have seen a recent revival for performing quantum chemistry simulations, especially for treating strongly correlated systems~\cite{tubman2016-1,holmes2016,evan2016,evangelista2014,wenjian2016}. 
Much of the recent interest in selected CI was sparked by the demonstration that the adaptive sampling CI method (ASCI) can attain an accuracy comparable to the density matrix renormalization group (DMRG) for Cr$_{2}$ with relatively little computational cost~\cite{tubman2016-1,holmes2016}. 
Selected CI techniques can be applied to a wide range of atomic, molecular, and solid state chemical systems. As such, they are a fundamentally different from DMRG, which is often the method of choice for low dimensional solid state systems.

Notable selected CI approaches developed in the last few years include ASCI~\cite{tubman2016-1} and the similarly titled adaptive CI approach (ACI)~\cite{evan2016}. An integral driven search extension to ASCI, known as heat bath CI (HBCI), was also recently developed~\cite{holmes2016}.  Additionally the classic configuration interaction perturbatively selected iteratively (CIPSI) algorithm has also seen recent advances~\cite{huron1973,harrison1991,scemama2013,giner2013,evan2016}. 
The development of the ASCI method~\cite{tubman2016-1} also demonstrated a connection between selected CI and the full configuration interaction quantum Monte Carlo (FCIQMC) technique ~\cite{booth2009,shepherd2012,kolodrubetz2013,hande2014,booth2013,daday2012,thomas2015,cleland2012,blunt2015}. 
 In contrast to FCIQMC however, the ASCI method utilizes a significantly more computationally efficient deterministic approach for searching Hilbert space, allowing it to produce nearly identical results at significantly reduced cost. Overall, the ASCI method is complementary to and competitive with density matrix renormalization group (DMRG)\cite{tubman2016-1} with regards to accuracy and computational cost. 
For one-dimensional systems DMRG is the method of choice due to the low entanglement of the wave functions~\cite{chan2002,white1999,white1993,amaya2015,hachmann2007,mizukami2013,sharma2014}.
However, DMRG is not an optimal choice when the entanglement is large, and virtually all chemically relevant systems have large entanglement~\cite{tubman2012,tubman2013,tubman2013-1,tubman2014,tubman2015,amaya2015}. This makes ASCI superior for many systems which include 2D and 3D systems, 
as well as for systems with large basis sets and for simulations of excited states~\cite{stoudenmire2012,schollwock2005}.

The ASCI algorithm of ref.~\cite{tubman2016-1} improves upon the perturbative approach of the CIPSI algorithm~\cite{huron1973,buenker1974,evangelisti1983,bagus1991,harrison1991,giner2013,caffarel2015,maynau2011} by introducing approximate search algorithms for finding the important determinants. 
In the original application of the ASCI algorithm, simulations of up to 48 electrons in 42 orbitals were made~\cite{tubman2016-1,Lehotola2017}. This work describes advances in the methodology that permit simulations with hundreds of orbitals on single workstations.
Few other methods are known to provide this level of speed and accuracy. Auxiliary field quantum Monte Carlo (AFQMC) is a competitive method, but it has several limitations that prevent consistent production of chemically accurate numbers~\cite{zhang2003,wirawan2015,chang2016}.  However there is always room to improve AFQMC type techniques with better trial wave functions~\cite{borda2018}.

The algorithms developed in this work are fundamentally driven by sorting based algorithms 
that are relevant to all current selected CI approaches. Our approach addresses the critical issue that selected CI methods are largely based on manipulation of large amounts of data.  As such, selected CI methods need to be designed to efficiently process and move data to and from the CPU.  Many of the improvements presented here make use of optimized tools and libraries that have not yet made their way into the selected CI literature.  We demonstrate here all these newly designed algorithms as extensions to the ASCI formalism. 
This includes algorithms that allow for different levels of parallelization and computing architectures, such as GPUs.  With these ideas, we are able optimize the ASCI algorithm beyond what has been done with previous selected CI approaches.

%%%%%%%%%%%%%%%%%%%%%%
\begin{table*}[htb!]
\centering
\begin{tabular}{|c|c|c|c|c|c|c|c|}
\hline
System&Dets&Basis&Main ASCI(s)& PT2(s) & Total time & ASCI Energy (Ha) & Ref. Energy (Ha) \\
\hline
C$_{2}(12e,28o)$&10$^{4}$&cc-pVDZ& 7& 4&11& -75.731895 &-75.731958~\cite{amaya2015}\\
C$_{2}(12e,28o)$&10$^{5}$&cc-pVDZ& 110 & 44 & 154 &-75.731954 &\\
C$_{2}(12e,28o)$&10$^{6}$&cc-pVDZ& 3740&570&4310&-75.731962&\\
\hline
C$_{2}(12e,60o)$&10$^{4}$&cc-pVTZ& 45 &70 & 115&-75.808698 &-75.809285~\cite{amaya2015}\\
C$_{2}(12e,60o)$&10$^{5}$&cc-pVTZ& 360& 600 & 960&-75.809190 &\\
\hline
C$_{2}(12e,110o)$&10$^{4}$&cc-pVQZ& 300 & 401 & 701 & -75.856781 &-75.85728~\cite{amaya2015} \\
C$_{2}(12e,110o)$&10$^{5}$&cc-pVQZ& 1020 & 4290 &5310 & -75.856822 & \\
\hline
Cr$_{2}$(24e,30o)&10$^{4}$&SVP& 18 & 11 & 29 & -2086.417192 &-2086.420948~\cite{amaya2015} \\
Cr$_{2}$(24e,30o)&10$^{5}$&SVP& 100 & 112 & 212 & -2086.419546 & \\
Cr$_{2}$(24e,30o)&10$^{6}$&SVP& 1250 & 1111 & 2361 & -2086.420438 & \\
Cr$_{2}$(24e,30o)&2*10$^{6}$&SVP& 2680 & 2215 & 4895 & -2086.420517 & \\
\hline
\end{tabular} 
\caption{An example demonstration of ASCI calculations performed in this work. All timings are in seconds. The basic ASCI algorithm includes an initial SCF calculation (Hartree-Fock), ASCI Search, diagonalization, orbital rotations. It includes extra diagonalizations that we perform after the orbital rotations, as well as refinement search and diagonalization steps after we finish growing the ASCI wave function. Many of these additional steps are included to make sure that we have a high quality wave function for benchmarking purposes. The orbital rotations and the extra diagonalization steps are not necessary to converge most of the simulations presented in the results section to chemical accuracy. See Table~\ref{tab:hbcomp} for direct comparisons to the HBCI algorithm. The calculations presented here were performed on a single core of a  Intel Xeon CPU E5-2620 2.1 GHz.}
\label{tab:timings}
\end{table*}%
%%%%%%%%%%%%%%%%%%%%%%

The different techniques presented in this work are outlined in the sections designated below.
\begin{itemize}
\item Constructing the Hamiltonian and density matrices (Section~\ref{sec:hamcons})
\item Search and pruning based on fast sorting algorithms (Section~\ref{sec:searchmain})
\item Other algorithm improvements (Section~\ref{sec:misc})
\subitem Fast diagonal matrix elements (Section~\ref{sec:fdiag})
\subitem Informed bit string representations (Electron representation and Difference representation) (Section~\ref{sec:ibit}) 
\subitem Hashed bit string representations (Section~\ref{sec:hbit})
%\subitem Efficient Scheduling (Section~\ref{sec:sched})
%\subitem GPU search algorithms (Section~\ref{sec:searchmain})
\end{itemize}
After we present these techniques, we apply our algorithm to 55 benchmark molecular systems in the G1 set (Section~\ref{sec:results}). Before presenting the details of the new algorithms however, we first present some timings and accuracy results to demonstrate their capabilities.

\begin{table*}

\centering
\begin{tabular}{|c|c|c|c|c|c|c|c|}
\hline
Comparisons&Dets&Basis&  Time PT2 & E(Variational) &E(HBCI+PT2) & E(Exact PT2) \\\hline
C$_{2}$(8e,58o) HBCI~\cite{sharma2017} & 142467 & cc-pVTZ &  80 & -75.7738 & -75.7846(3)&\\
C$_{2}$(8e,58o) ASCI & 50000 & cc-pVTZ &  60 & -75.768939 & &-75.784113 \\
C$_{2}$(8e,58o) ASCI & 100000 & cc-pVTZ & 117 & -75.773560 && -75.784468 \\
C$_{2}$(8e,58o) ASCI & 142467 & cc-pVTZ & 166 & -75.775386 & &-75.784589 \\
\hline
F$_{2}$(14e,58o) HBCI~\cite{sharma2017} & 395744 & cc-pVTZ &  120 & -199.2782 & -199.2984(9)&\\
F$_{2}$(14e,58o) ASCI & 20000 & cc-pVTZ & 60 & -199.254301 && -199.295491 \\
F$_{2}$(14e,58o) ASCI & 100000 & cc-pVTZ & 300 & -199.270670 && -199.296289 \\
F$_{2}$(14e,58o) ASCI & 300000 & cc-pVTZ & 891 & -199.278140 && -199.296686 \\
F$_{2}$(14e,58o) ASCI & 395744 & cc-pVTZ  & 1163 & -199.279209 & &-199.296767 \\\hline
\end{tabular}
\caption{A comparison of HBCI to ASCI ground state energies. HBCI results are taken from ref.~\cite{sharma2017}. HBCI uses a stochastic algorithm to perform the PT2 for these system sizes. The results here show that HBCI is inefficient in generating a variational wave function, and does not always give accurate PT2 energies, due to its large error bars. In simulations where ASCI and HBCI use the same variational wave function, the deterministic ASCI results will be always be arbitrarily more accurate due to lack of stochastic error. The HBCI results were performed on nodes with 2 Intel Xeon E5-2680 v2 processors of 2.80 GHz.  These processors have 20 computational cores per node.   The ASCI simulations were performed on a single core of a Intel Xeon E5-2620 v5 processor of 2.10 GHz. We have calculated the equivalent single core time in order to make a comparison to our calculations and we have scaled the ASCI timings to be representative of a 2.8 GHz core. For the PT2 simulations, we truncate the contributions less than 10$^{-8}$, as in Ref.~\cite{sharma2017}. Orbital rotations and refinement steps were turned off for these ASCI simulations,  as these extra steps are not present in HBCI. For a given number of determinants, the ASCI wave functions  is better than the HBCI wave functions (as expected), even without the extra steps.  An HBCI simulation with stochastic PT2 accurate to 0.1 mHa (with 68\% probability) would be 6 and 11 times slower than the ASCI timings presented here for C$_{2}$($N_{dets}$=100000) and F$_{2}$($N_{dets}$=300000), respectively.  However, stochastic HBCI calculations with 0.1 mHa accuracy (with 95\% probability) for calculating energy differences between two calculations (like an atomization energy), would be 48 and 88 times slower than ASCI. }
\label{tab:hbcomp}
\end{table*}%

\section{New Algorithms and Timings}
\label{sec:algotime}

\begin{figure}
\begin{center}
\scalebox{1}{\includegraphics[width=1.0\columnwidth]{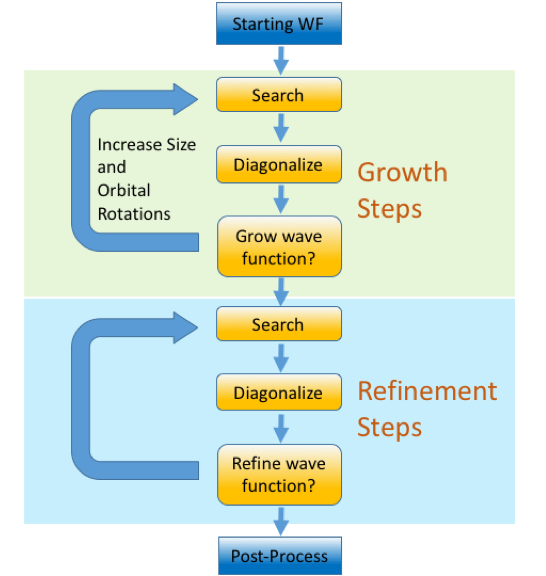}}
\end{center}
\caption{A flowchart of the ASCI algorithm. The main computational parts are the search, diagonalization, and post processing steps. 
The growth steps are done in the first set of iterations of ASCI to bring the variational wave function from the Hartree-Fock determinant to a wave function of size $N_{tdets}$. We grow the wave function, since we find it is slower to perform diagonalizations on a full size but inaccurate wave function.  We use the refinement steps when we want to generate a very high accurate variational wave function. During the refinement step we fix the size of the wave function but continue to improve upon it through search/diagonalization iterations. }
\label{fig:flowchart}
\end{figure}
Figure \ref{fig:flowchart} presents an overview of the ASCI algorithm.  The main components of the algorithm involve building the Hamiltonian, diagonalization, search, and second order perturbation theory (PT2).  We break up the wave function construction into a growth and a refinement process. The growth process involves several steps in which the wave function is increased in size and the orbitals are rotated. The refinement process is a set of final steps in which the size of the wave function is kept fixed, but the quality is improved.  
In the following sections we introduce new algorithms for ASCI, which include approaches for search and Hamiltonian construction.  We present several algorithm choices for these components.  Our current recommendation is to use the following algorithms for constructing a selected CI: \textit{New ASCI search} (described in this work), \textit{dynamic bit masking} for Hamiltonian construction (described in this work), \textit{fast orbital rotations} (described in this work) and a \textit{deterministic PT2} based on sorting, which will be described in a future work. We also introduce the residue arrays algorithm for Hamiltonian construction, which may be more efficient than dynamic bit masking for certain Hamiltonians. In particular, the size of a residue array is determined by the number of electrons being simulated, thus a residue array is more  efficient in the limit in which a system has a small number of electrons.
 To illustrate the current set of algorithms, and to compare them to previously published work, we present a series of timings in Tables~\ref{tab:timings}, \ref{tab:hbcomp}, and \ref{tab:comparisons}.

In Table~\ref{tab:timings}, we present ASCI timing results for C$_{2}$ and Cr$_{2}$. These timings represent a careful simulation of the variational wave function that is designed to generate highly accurate results. More aggressive approximations can lead to much faster calculation of the variational wave function. In Table~\ref{tab:hbcomp} we demonstrate the timings for such an aggressive approach in order to make a comparison to HBCI.  The timings for the PT2 simulations presented here represent the fastest published timings for deterministic algorithms that we are aware of. Indirect comparisons to stochastic PT2 in Table~\ref{tab:hbcomp} demonstrate that stochastic algorithms are only competitive if large stochastic errors are acceptable. Recent papers~\cite{sharma2017} have suggested that stochastic PT2 is the best way to efficiently simulate large system with selected CI. However, our comparisons to the stochastic method, indicate that it is unclear 
%for exactly 
when this would be true and for which ranges of accuracy.  In fact, the errors on stochastic methods get even larger when energy differences are taken, which is important for chemically relevant results.  The details of the sorting based deterministic perturbation theory algorithms will be published in a future work, and we hope in that work to provide a detailed analysis to compare with stochastic approaches to perturbation theory. 

\begin{table}[htb!]

\centering
\begin{tabular}{|c|c|c|c|c|c|c|c|}
\hline
Comparisons&Dets&Basis& Time(s)& E+PT2(Ha) \\
\hline
old ASCI Cr$_{2}$(24e,30o)~\cite{tubman2016-1}&10$^{4}$&SVP& 1000 &-2086.4177 \\
old ASCI Cr$_{2}$(24e,30o)~\cite{tubman2016-1}&10$^{6}$&SVP& 133000 &-2086.4203 \\
\hline
CIPSI Cr$_{2}$(24e,30o)~\cite{quantumpackage}&4*$10^{4}$&SVP& 4290 &-2086.41818 \\
ASCI Cr$_{2}$(24e,30o)&4*10$^{4}$&SVP& 82  &-2086.41808 \\
HBCI~\cite{holmes2016} Cr$_{2}$(24,30)&4*10$^{4}$&SVP& 120 &-2086.42130 \\
%ASCI (this work, inaccurate) Cr$_{2}$(24,30)&4*10$^{4}$&SVP& 35 & 45 & 70 &-2086.42099 \\
ASCI (fast) Cr$_{2}$(24e,30o)&4*10$^{4}$&SVP& 55 &-2086.42099 \\
\hline
ASCI C$_{2}$(12e,60o)&4*10$^{4}$&cc-pVTZ& 280&-75.80898 \\
HBCI~\cite{holmes2016} C$_{2}$(12e,60o)&4*10$^{4}$ &cc-pVTZ& 540 &-75.80873 \\

\hline
\end{tabular}
\caption{A comparison of the current work with the original ASCI results and other selected CI methods. In the original ASCI paper (labled 'old ASCI'), slow but memory-efficient algorithms were used for the PT2. The Cr$_{2}$ comparisons in the middle columns demonstrate the difference between accurate wave functions (CIPSI and ASCI) and wave functions in which a minimal search algorithm is used such as in HBCI, and 'ASCI (fast)'. For 'ASCI (fast)' the search parameters are turned down significantly so as to be similar to HBCI. Although the energies for HBCI and 'ASCI (fast)' are close to the converged DMRG results, this is because, for these inaccurate wave functions, the energy is converging from below. The HBCI energy for Cr$_{2}$ will observe significant non-monotonic behavior before converging. The timings for the different algorithms were performed on different computing architectures.  The CIPSI simulation was performed on a single core of an Intel Xeon E5-2680v3 processors of 2.5 GHz.  The ASCI simulations were performed on a single core of an Intel Xeon E5-2620 v5 processor of 2.10 GHz.  The HBCI timings were taken directly from ref.~\cite{holmes2016}.   The timings are meant only to give a semi-quantitative comparison, as the timings for the ASCI includes many extra steps which are not included in the CIPSI and HBCI timings, as discussed in the main text.}
\label{tab:comparisons}
\end{table}%

The timings for ASCI in Table~\ref{tab:timings} include the initial SCF calculations, orbital rotations, and extra diagonalization steps that are required after performing an orbital rotation. Additionally after the growth of the wave function we perform several refinement iterations, which includes both search and diagonalization steps, to improve the wave function as much as possible before the PT2. The column labeled `Main ASCI' incorporates all of these different steps. When using the new ASCI algorithms for Hamiltonian building, orbital rotations, and search, the Hamiltonian diagonalization is now the dominant bottleneck in the limit of large determinants. Performing an accurate search step can be done quickly with the new ASCI search algorithms, and thus is no longer the bottleneck of a simulation. 

 In Table~\ref{tab:comparisons} we present a comparison to selected CI calculations from previous published works, as well as CIPSI simulations as implemented in the Quantum Package suite for the Cr$_2$ dimer~\cite{quantumpackage}.
 We present a comparison to our original ASCI implementation 'old ASCI' (the details of which can be found in Ref.~\cite{tubman2016-1}), which uses a memory efficient but computationally slow PT2 correction. The next sets of results are generated with 'HBCI' and 'ASCI (fast)', which use a less accurate, but faster, search algorithm than ASCI and CIPSI.  These results illustrate problems that arise from using an inaccurate search algorithm and generating medium quality variational wave functions.  We observe in some situations that low quality wave functions  can result in significant non-variational behavior for the perturbation theory correction. 
 Unconverged wave functions can also lead to slow energy convergence from above, as can be seen in comparing HBCI to ASCI results for the C$_2$ dimer in Table~\ref{tab:comparisons}.

In Table~\ref{tab:hbcomp}, we remove the orbital rotations and extra diagonalizaton steps after the growth algorithm in order to make more direct comparisons (in terms of computational timing) to HBCI. In these comparisons HBCI perturbation theory results are generated using a stochastic algorithm in order to treat large basis sets. This is not necessary for ASCI, and all the ASCI simulations presented in this work are completely deterministic. The results presented here show that ASCI generates lower energy variational wave functions than HBCI for the same number of determinants.
For example, the C$_{2}$ ASCI($N_{dets}$=100,000) and F$_{2}$ ASCI($N_{dets}$=300,000) give comparable variational energies to HBCI but with 42,000 and 95,000 fewer determinants, respectively. This means that the HBCI wave function is missing many important determinants, as evidenced by the more compact ASCI wave function. 

It needs to be emphasized that the PT2 comparisons have to be done with care, since the stochastic HBCI results have both systematic bias and stochastic bias, whereas the ASCI algorithm presented here has no stochastic errors. For timing comparisons of the PT2 results, we will consider the C$_{2}$ ASCI($N_{dets}$=100,000) and F$_{2}$ ASCI($N_{dets}$=300,000) simulations because they have nearly the same variational energy as the HBCI results. The error bar for stochastic perturbation theory decreases like $\frac{1}{\sqrt{N_{samp}}}$, where $N_{samp}$ is the number of samples for computing the stochastic average.  For an accuracy of 0.1 mHa, the C$_{2}$ and F$_{2}$ ASCI perturbation theory timings would be 6 times faster and 11 times faster than HBCI, respectively.  This is for a 68\% likelyhood that the result is within 0.1 mHa of the actual answer.  A more definitive result with a 95\% likelihood would be would be 24 and 44 times faster with ASCI than HBCI.  To calculate energy differences, such as for calculating atomization energies, another factor of two is needed to make sure the energy difference is 0.1 mHa accurate, for which ASCI would be 48 and 88 times faster than HBCI for these two systems respectively.   In these scenarios, the ASCI results would still be more accurate than HBCI, as the ASCI results are calculated without stochastic errors. For an accuracy of 0.01 mHa, the ASCI timings presented are several hundred times faster than the HBCI. 

In the case of C$_{2}$, the fully converged ASCI simulation, which we perform with orbital rotations, has an energy of 75.78508(Ha). The results in Table~\ref{tab:comparisons} show slow convergence because these results are calculated with Hartree-Fock orbitals. In comparing to ASCI, these results show how both systematic errors and stochastic errors play a role in the HBCI result. The HBCI F$_{2}$ results are significantly non-variational, which is likely due to the large stochastic errors, coupled with a variational wave function that is of only medium quality. We also note that for all ASCI simulations in this work, other than those in Table~\ref{tab:hbcomp}, orbital rotations are always included and significantly improve the convergence. Orbital rotations allow ASCI to generate a more compact wave function with fewer determinants.  The calculation of the 1-RDM used in orbital rotations in ASCI are calculated for almost no cost (described in Section~\ref{sec:fdiag}). 

We also note here that comparable DMRG simulations use a significant amount of CPU time and memory compared to the results presented here.   The largest DMRG calculations in ref.~\cite{amaya2015} (butadiene: 22e,88o) reportedly used 850 GB of memory, 17 TB of disk space, and over a 1000 CPU hours per sweep with a bond dimension of 3000.  Although similar sized systems (and even larger) are presented in this work, we note that DMRG is variational whereas ASCI+PT2 is not.  Regardless, both methods produce high level results and can be used to provide validation of each other as they use very different approaches to solving quantum Hamiltonians.

%%%%%%%%%%%%%%%%%%%%%%
%%%%%%%%%%%%%%%%%%%%%%
\section{Theoretical Background: selected CI algorithms}
\label{sec:theory}
%%%%%%%%%%%%%%%%%%%%%%

The main idea behind the selected CI approach is to perform diagonalization on a determinant space in which one captures as many important degrees of freedom as possible. This is the principle behind all exact diagonalization and CI techniques, although most methods do not allow for explicit searching for important determinants~\cite{lauchli2011,vogiatzis2017,szabo:book,gan2006,gan2005,sherrill1999,abrams2004,szalay2012,bender1969,buenker1978,roth2009,bagus1991}. 
Thus in contrast to more traditional CI techniques, the idea of using a selected CI approach is to generate determinants that account for 90\% or more of the top contributions to the full CI wave function.

In the ASCI method, as well as in most selected CI methods, a wave function, $\psi_{k}$, is iteratively improved  to reach a desired accuracy. Here we use $k$ to index the current iteration of the algorithm.  The search part of the algorithm requires two rules: a selection criterion to determine what part of Hilbert space to search from (pruning) and a ranking criterion to determine the best determinants to include in the improved wave function $\psi_{k+1}$. For the algorithms considered here the ranking criterion is derived from a consistency relationship among the coefficients of eigenstates of the Schr\"{o}dinger equation. The consistency equation is given as follows if we consider an expansion of an eigenstate in terms of its coefficients $C_i$:
\begin{equation}
C_{i} = \frac{\sum_{j \ne i}H_{ij}C_{j}}{(E-H_{ii})}
\end{equation} 
\noindent where $H_{ij}$ is the Hamiltonian matrix element between the $i$th and $j$th determinant, and $E$ is the energy of the eigenstate.
If we reinterpret this equation, we can use it to predict a new and better set of determinants for expanding $\psi_{k+1}$ by taking the LHS as an estimate of the magnitude of the expansion coefficients, % using the expansion coefficients of wave function $\psi_k$,

\begin{equation}
A_{i} = \frac{\sum_{j \ne i}H_{ij}C^k_{j}}{(E_k-H_{ii})},
\label{eqn:rankeqn}
\end{equation}

\noindent where $C^k_j$ is the CI expansion coefficient of the $j$th determinant and $E_k$ is the energy of the wave function in the $k$th iteration. It is useful to think of $C^{k}_i$ as the coefficients of an input wave function, and the output coefficients, $A_{i}$, as an estimate of coefficients of an improved wave function (close to the ground state). Since the goal of selected CI is to include the most important weight determinants in the expansion, we use this equation to define a ranking, where $A_i$ is the rank value of the $i$th determinant. These $A_i$ coefficients are related to a first-order perturbation estimate for CI coefficients in many body perturbation theory~\cite{huron1973}. 

 In practice this iterative approach generates all the top contributions to the wave function. Having the top contributions is critical to obtain highly accurate energies, as was recently shown with the ASCI method in combination with second order many body perturbation theory~\cite{tubman2016-1}. 
The perturbation theory energy is calculated with Epstein-Nesbet PT2 which is given as, 
 \begin{equation}
 E_{PT2} = \sum_{i}\frac{|\langle\psi|H|D_{i}\rangle|^{2}}{H_{ii} -E_{}}. \label{eqn:en}
 \end{equation}

To facilitate the following discussion, we introduce the following notation: $\psi_k$ is the wave function at the $k$th iteration of the selected CI algorithm, $C_i$ is the coefficient of the $i$th determinant in the expansion of $\psi_k$, $D_i$ is the $i$th determinant in $\psi_k$, $\{D\}$ is the set of all determinants in $\psi_k$, \{$D_{i}^{sd}$\} is the set of all determinant which are single or double excitations from determinant $i$. In general, $i\in \{D\}$, and \{$D^{sd}$\} is the set of all determinants which are single and double excitations from $\{D\}$ (that is \{$D^{sd}$\}=$\bigcup_i$ \{$D_i^{sd}$\}). 
We describe two determinants which differ by a single or double excitation as \textit{connected}. 
The notation we use here is summarized in Table~\ref{tab:symbols}.

\begin{table}[]
\centering
\begin{tabular}{lp{60mm}}
\hline
\multicolumn{1}{|l|}{Symbol} & \multicolumn{1}{l|}{Explanation} \\ \hline
$\psi_k$ & The wave function in the current ASCI step\\
$C_i$ & The coefficients of the $i$th deteminant $D_i$ of $\psi_k$ \\
$D_{i}$ & The $i$th determinant in $\psi_k$ \\
\{$C$\} & The set of coefficients in $\psi_k$ \\
$\{D\}$ & The set of determinants in $\psi_k$ \\
\{$D^{sd}$\}& The set of all single and double excitations that are connected to $\psi_k$ \\
\{$D_{i}^{sd}$\} & The set of all single and double excitations that are connected to the determinant $D_{i}$ \\
%\{$D_{connect}^{sd}$\} & The set of all number pairs ($i$,$j$) such that $i$ is a determinant in $\{D\}$ and for a given $i$, the $j$ index must be in~\{$D_{i}^{sd}$\}. \\
$N_{tdets}$ & Number of determinants in the current wave function. \\
$N_{cdets}$ & Core space size, used for pruning in ASCI\\
\end{tabular}
\caption{A reference to a list of the symbols used in this work. }
\label{tab:symbols}
\end{table}

\subsubsection{Introduction to the Computational Challenge }
\label{sec:cc}

The modern implementations to exact diagonalization and selected CI described in this work are in part focused on calculating Eq.~\ref{eqn:rankeqn} and Eq.~\ref{eqn:en} efficiently. The main issue for using these equations is that they involve all the elements in \{$D^{sd}$\}, which can be extremely large. For general systems of interest a set size of 10$^{12}$ and larger is not uncommon. Furthermore, for each element $i$ in \{$D^{sd}$\}, all ways in which the Hamiltonian connects from $\{D\}$ to $i$ must be found. This can becomputationally expensive, especially for wave functions that can have between 10$^{5}$ to 10$^{9}$ determinants (although recent full CI simulations have recently gone all the way up to 10$^{12}$ elements~\cite{vogiatzis2017}). 
The straightforward and slow  approach to performing this calculation is as follows: for all $i$ in $\{D\}$ generate the set \{$D_{i}^{sd}$\}. Then, for each element $a$ in \{$D_{i}^{sd}$\}, search $\{D\}$ to find all the elements $b$ such that $H_{ab}$ is non-zero (that is $b$ is singly or doubly connected to $a$). This approach scales as $\mathcal{O}((N_{tdets}*N_{occ}*N_{virt})^{2})$, where $N_{occ}$ is the number of occupied orbitals and $N_{virt}$ is the number of virtual orbitals. The asymptotic scaling for $N_{tdets}$ (that is, for a fixed number of orbitals we ignore $N_{occ}$ and $N_{virt}$) can be seen as follows: the number of terms in \{$D^{sd}$\} is proportional to $N_{tdets}$, and, for each term, a search over the $N_{tdets}$ determinants of the wave function is needed to find all connections. Constructing the Hamiltonian also requires a similar algorithm, where connections among the elements of $\{D\}$ need to be found. Improvements to this problem has been studied previously with the introduction of residue trees~\cite{stampfub2005}.  For both the search and Hamiltonian construction, we present new algorithms in the modern implementation sections.

\begin{figure}[tpb]
\begin{algorithm}[H]
 \begin{algorithmic}[1]
\State Input $\psi_k$ and an initially empty array \{$V$\} that will hold information in pairs of (determinant bit string, ranking value)
\State Generate $D^{sd}$ from $\psi_{k}$. 
\begin{itemize}
\item Iterate through $i \in \{D_{}\}$, and find \{$D_{i}^{sd}$\}
\item If CIPSI search, generate all \{$D_{i}^{sd}$\}
\item If original ASCI search, prune terms in \{$D_{i}^{sd}$\} based on \{C$_{}$\}
\item If HB search, prune terms in \{$D_{i}^{sd}$\} based on \{$C_{}$\} and $H_{ij}$
\end{itemize}
\State Calculate a ranking either approximately or exactly from Eq. \ref{eqn:rankeqn}.
\State Gather all the top contributing determinants to determine $\psi_{k+1}$ for use in the next iteration 
 \end{algorithmic} 
 \caption{Generic Search (pruning and ranking)}
 \label{alg:genericsearch}
\end{algorithm}
\end{figure}
\subsubsection{CIPSI}
\label{sec:CIPSI}
A straightforward idea for searching for important determinants is to generate all terms in $\{D^{sd}\}$, rank them by Eq.~\ref{eqn:rankeqn}, and take the top $N_{tdets}$ of them to form the basis for creating $\psi_{k+1}$. This is the approach used in CIPSI and is defined by a  single parameter $N_{tdets}$ that determines the number of determinants that are retained in a simulation.  No effort is made to prune the search space in such an approach. As demonstrated explicitly in the ASCI approach, this pruning is important for computational efficiency and results in no significant loss in accuracy. For most CIPSI applications the generation of the search space is the most costly step of the calculation. Pruning the search space and approximating the ranking algorithm become critical for going to larger system sizes. 

\subsubsection{ASCI}
\label{sec:ASCI}

The ASCI algorithm, introduced in ref.~\cite{tubman2016-1}, allows selected CI approaches to be more computationally efficient in the search algorithm. In ASCI, only a subset of \{$D^{sd}$\} are considered for ranking, resulting in a significant increase in efficiency. This is made possible by having a good criterion (using the structure of the wave function) for quickly determining which sets $\{D_i^{sd}\}$ are likely to have highly ranked connections, allowing one to avoid searching unimportant parts of Hilbert space. %ASCI achieves this by introducing a simple pruning algorithm so that the ranking algorithm is calculated only on a limited number of terms.

The ASCI pruning algorithm uses the magnitude of the \{$C_{}$\}. %before calculating Eq.~\ref{eqn:rankeqn} as the ranking criterion. 
Only connections from the top $N_{cdets}$ determinants in $\psi_k$ (where these are ranked by $C_i$ value) are considered. Each ASCI algorithm iteration is parametrized by two determinant subspaces: a \textit{core space} of size $N_{cdets}$ and a \textit{target space} of size $N_{tdets}$. This leads to an iterative algorithm with significantly more efficient performance than CIPSI methodologies.

\subsubsection{Integral driven search extension of ASCI (Heat Bath)}
\label{sec:heatbath}

The speed of the search component of ASCI may be accelerated by sacrificing some accuracy. In the HBCI approach, a new pruning criterion was introduced, using a combination of the wave function coefficients and the Hamiltonian matrix elements. This can be understood as an integral driven search approach in the ASCI formalism. The vast majority of connections in the search step are due to double excitations, whose matrix elements are just the antisymmetrized two-electron integrals, which can be sorted once at the beginning of the algorithm run. The HBCI pruning criterion is as follows: all single excitation and only those double excitations such that $c_iH_{ij}>\epsilon$ are selected for ranking, where $\epsilon$ is a parameter that replaces $N_{cdets}$ to define the integral driven search algorithm. 

In the original heat bath approach~\cite{holmes2016}, the ranking criterion was replaced by $A_j=max_{j}(H_{ij}C_{i})$ for the doubles contributions, which essentially allows for the pruning and ranking to occur simulatenously. With this modified ranking criterion, both the denominator and the phase information are ignored. This approximation is less justifiable for more difficult quantum chemistry problems. As shown in Table~\ref{tab:comparisons}, this reduces HBCI's ability to generate a compact wave function for C$_{2}$ and Cr$_{2}$. 
Additionally, as the basis set size is increased, the energy denominator can become large and many determinants can become unimportant due to the involvement of high energy orbitals. Thus, the denominator in Eq.~\ref{eqn:rankeqn} can become increasingly important for pruning Hilbert space. For ASCI, we find that the new search algorithms are sufficiently fast that we do not need to use the approximations in the HBCI formalism. Nevertheless, integral driven search algorithms are compatible with the sorting algorithms presented in this work. % and can be more efficient for some systems.

\section{Modern Implementation}
\label{sec:algtechs}
In this section, we present the newest algorithms for ASCI. The selected CI algorithms presented here are extremely efficient on modern computing architectures, with added functionality that allows for scaling to large numbers of electrons and basis sets. %These algorithms definitively improve upon all published selected CI algorithms that we are aware. 
This section is organized by first presenting several different algorithms for the critical ASCI functions of determinant search and Hamiltonian build, and then describing how the best of these algorithms can be efficiently implemented. 
The main parts of the algorithm that we describe here are the \textit{Hamiltonian construction step}, presented in Section \ref{sec:hamcons}, and the \textit{search algorithm}, presented in Section \ref{sec:searchmain}. We also briefly discuss the \textit{perturbation theory step}, but the optimal algorithmic implementations of this will be presented in a future work.

\subsection{Sorting as a paradigm for selected CI}

In this section we present evidence for the efficiency of using sorting based algorithms in selected CI on modern computers. We start by first considering both a generic 'search algorithm' and the 'Epstein-Nesbet perturbation theory' algorithm in the limit of unlimited memory. The following results were developed through extensive testing of various libraries such as the standard template library (STL)~\cite{plauger2000} and Boost~\cite{boost2011}.  We specifically considered different solutions that can be developed through sorting, hash tables, and search trees. 
 
The idealized unlimited memory algorithm is as follows: Generate \{$D^{sd}$\} by going through each $i$ in \{$D_{}$\} and retain both $i$ and the connected bit string $j\in \{D^{sd}\}$ in an array. Sort the array based on $j$ and now the elements will be grouped together such that that calculation of Eq.~\ref{eqn:rankeqn} or Eq.~\ref{eqn:en} can be done with a single pass over the sorted list. We hypothesize that there is no faster way to perform this process than by using a sorting algorithm on modern computing architectures.  However, very recent advances with hash tables suggest that such structures might eventually also become competitive, particularly for large basis sets~\cite{hash_benchmark}.  For realistic algorithms with limited memory, the sorting algorithm can be modified in different ways. We present the 
most efficient new ASCI search algorithm using this approach in section \ref{sec:searchmain}.

\subsubsection{The world of sorting}
\label{sec:wws}
The dominant bottleneck in many selected CI algorithms is cache inefficiency, that is, an inability to get all of the necessary data to the CPU in a way that computation can occur in an efficient manner. Practical sorting algorithms are developed to be as cache efficient as possible and are thus a natural choice for ASCI. Research on sorting algorithms is an active field with  new algorithms being developed to work with modern computing architectures~\cite{radulsort,pdq,vergesort,skasort,musser1997,timsort}. There have been many significant innovations even within the last few years~\cite{2017arXiv170408579B,blocksort}. Additionally, parallelization of sorting algorithms is quite different on GPUs versus CPUs, and efficient parallelization approaches are only starting to be developed~\cite{ips4,2015arXiv151103404B}. 
 
 Since the selected CI algorithms presented in this work are based on sorting; any developments made in improving sorting algorithms also improves these algorithms. We emphasize that sorting based algorithms allow for easy access to parallelization which include GPU implementations. To demonstrate the efficiency of different sorting algorithms, we present a comparison of different methods in Table~\ref{tab:speedsort}. The STL implementation is a quicksort and PDQ is a pattern defeating quicksort. The Boost sorting algorithm is spreadsort.   Spreadsort is a hybrid algorithm that uses a radix sort in most situations. The IPS$^4$O algorithm is a sorting algorithm designed to be run in parallel. For our small tests,  we found the algorithm to be nearly linear scaling up to 8 cores. Our GPU tests were performed with the Thrust library~\cite{thrust} and the timing results include the time is takes to move the data on and off the GPU.

\begin{table}
\centering
\begin{tabular}{|c|c|}
\hline
Algorithm & Timing(s) \\
\hline
(STL) Quick Sort~\cite{plauger2000} & 30\\
PDQ Sort~\cite{pdq} & 29\\
(Boost) Spreadsort~\cite{boost2011} & 31\\
IPS$^4$O 1-core~\cite{ips4} & 37\\
IPS$^4$O 4-cores~\cite{ips4} & 9\\
IPS$^4$O 8-cores~\cite{ips4} & 5\\
(Thrust) GPU sort~\cite{thrust}& 2.3\\
\hline
\end{tabular}
\caption{Comparison of different sorting techniques over 128 bit determinants from a Cr$_{2}$ SVP ASCI simulation. The test is for sorting 300 million integers. Most of the sorting algorithms presented here have similar performance. STL, Boost, and Thrust are popular libraries that include many application tools. The PDQ is a pattern defeating quick sort. The IPS4O and GPU results show that the sorting parts of the ASCI algorithm can be either parallelized or offloaded to a GPU for enhanced performance. The CPU simulations were performed on an Intel Xeon E5-2620 v5 processor of 2.10 GHz. The GPU calculation was performed on a NVIDIA Kepler K80.}
\label{tab:speedsort}
\end{table}%

\subsection{Hamiltonian construction}
\label{sec:hamcons}
Constructing the Hamiltonian requires some considerations beyond those required for traditional CI algorithms. At any given step of an iterative selected CI calculation, one has to determine which Hamiltonian matrix elements are non-zero~\cite{szabo:book}. In a typical active space calculation, all determinants within the active space are present and it is trivial to find the non-zero matrix elements. For a selected CI simulation this is not the case. %The main algorithm design in this section is to find the non-zero matrix elements of the Hamiltonian, given a list of determinants that define the Hilbert space for the Hamiltonian. 
The tests we present in this section take a list of determinants as input, and output a unique set of matrix coordinates that are the non-zero matrix elements.  These matrix elements must be ordered by row to be used in a sparse matrix diagonalization routine.

%The results of our tests are shown in figure~\ref{fig:speedhamil} and Table~\ref{tab:speed}.
 There are two straightforward approaches typically used in Hamiltonian construction that are however computationally inefficient in the limit of large numbers of determinants.  These algorithms are presented for completeness in algorithms~\ref{alg:doubleloop} and ~\ref{alg:singledouble}. The 'double loop' algorithm is fast for wave functions in which few determinants are being considered, since the number of operations scales as $O(N_{tdets}^{2})$. The 'singles/doubles' algorithm can be efficient when there are not many orbitals or electrons but a lot of determinants retained. In the limit of a large number of either determinants, electrons, or orbitals, these methods can become inefficient. 

\begin{figure}
\begin{algorithm}[H]
 \begin{algorithmic}[1]
\State Generate all determinant pairs in $\psi_k$
\State Determine if they are singly or double connected
\State If yes, calculate the matrix element 
 \end{algorithmic} 
 \caption{Hamiltonian Construction: Double Loop}
 \label{alg:doubleloop}
\end{algorithm}
\end{figure}

\begin{figure}
\begin{algorithm}[H]
 \begin{algorithmic}[1]
\State Generate $D^{sd}_i$, all singles and doubles from a determinant $D_i$ in current wave function $\psi_k$
\State For every determinant $j\in D_i^{sd}$, check if $j\in\{D\}$
\State For each match, calculate the matrix element
 \end{algorithmic} 
 \caption{Hamiltonian Construction: Singles and Doubles}
 \label{alg:singledouble}
\end{algorithm}
\end{figure}

\begin{figure}
\begin{algorithm}[H]
 \begin{algorithmic}[1]
\State For each determinant in $\psi_k$, create all bit strings/determinants in which two electrons are removed. Each one of these bit strings is called a residue.
\State Store the following two things together: (residue, list of all determinants that generate the residue) in a tree object \{$V$\}. Each time a new residue is found, query the tree for its existence. If it is new, add it to the tree. If the residue is already there, add the generating determinant to the residue's list.
\State Once the tree is finished, go through every residue. All pairs of determinants in a residue list are doubly connected, and have a non-zero Hamiltonian matrix element.
\State Go through the list of Hamiltonian matrix elements and remove any duplicates (Determinant pairs that are singly connected will appear in multiple residues with each other).
 \end{algorithmic} 
 \caption{Hamiltonian Construction: Residue Trees}
 \label{alg:residuetrees}
\end{algorithm}
\end{figure}

\subsubsection{Hamiltonian Construction: Residue Arrays}

An alternative approach to constructing the Hamiltonian uses a data structure called a residue tree. Residue trees are one of the fastest techniques currently in the literature for constructing Hamiltonians~\cite{stampfub2005}. The residue tree is a simple data structure that makes it straightforward to find connections between determinants. The \textit{residues} that can be generated from a reference determinant is a set of determinants that is generated by removing two electrons in all possible ways from the reference determinant. Each node of a residue tree contains a residue and a list of all the determinants in $\{D_{}\}$ that can generate the residue. The full residue tree consists of all the distinct residues that can be generated from $\{D_{}\}$. The residue tree is created as a tree so that any node in it can queried and found  in $\mathcal{O}(log(N))$ time (where $N$ is the number of nodes in the tree).  

The number of possible residues is approximately $N_{tdets} \binom{N_{elec}}{2}$, where $N_{elec}$ is the number of electrons in the system. In the case of Cr$_{2}$ with 48 electrons and $N_{tdets}$=300,000, the number of residues is on the order of 200 million. For our testing, the residue trees are implemented as red-black trees and related search trees through STL and Boost libraries. 
After experimenting with residue arrays we realized that the idea of a residue can be made even more efficient with a sorting-based algorithm on such arrays rather than a tree based algorithm.

Residue arrays are very similar to residue trees, but the tree structure is completely removed. Instead, each residue, together with the determinant that generated it, is stored in an unsorted array. After all residues have been generated, the array is sorted by residue. It is then in a form in which all non-zero matrix elements of the Hamiltonian can be generated, exactly as in the case of residue trees. Figure \ref{fig:speedhamil} and Table \ref{tab:speed} provide comparisons of residue arrays to other techniques for finding the non-zero Hamiltonian matrix elements. Figure \ref{fig:speedhamil} shows that in our Cr$_{2}$ ASCI test, residue arrays are second only to dynamic bit masking (discussed in Section \ref{sec:dynbit}).

\begin{figure}
\begin{algorithm}[H]
 \begin{algorithmic}[1]
\State For each determinant in $\psi_k$, create all bit strings/determinants in which two electrons are removed. Each one of these bit strings is called a residue.
\State Store the following two things together: (residue, determinant that generated residue) in an array \{$V$\}. 
\State Once all the residues have been generated, sort the array by residues. All residues that are equal will now be adjacent in the array. Pairs of determinants that generate a residue are doubly connected, and have a non-zero Hamiltonian matrix element.
\State Store all connections and then remove any duplicates (Determinant pairs that are singly connected will appear in multiple residues with each other).
 \end{algorithmic} 
 \caption{Hamiltonian Construction: Residue Arrays}
 \label{alg:residuearrays}
\end{algorithm}
\end{figure}

\subsubsection{Hamiltonian Construction: Dynamic Bit Masking}
\label{sec:dynbit}

\begin{figure}
\begin{algorithm}[H]
 \begin{algorithmic}[1]
\State Determine the number $n$ of bit masks to use, and set the bit masks. This is done by looking at the occupation of the orbitals given by $\psi_k$. The more orbitals that are close to being 50/50 occupied, the more efficient the dynamic bit masking becomes on increasing $n$. 
%\item Determine a set of $n$ bit masks according to the set of determinants to be diagonalized (dynamically selecting bit masks)
\State Apply each bit mask to each determinant in $\psi_k$, and save the masked value mask(j)\& detstring(i) = mv(i,j): i.e., determinant $i$ with mask $j$ applied to it.
\State Pick (n-4) of the bit masks and using the related mv(i,j), create a composite number out of them for each $i$. %Put this number along with the determinant bit string, put into a sortable object.
\State Find all pairs of determinants with the same composite number. These are non-zero matrix elements of the Hamiltonian (check for false positives). Repeat the last two steps for all $n \choose (n-4)$ possible ways of picking (n-4) bit masks.
\State Go through the list of Hamiltonian matrix elements and remove duplicates.
 \end{algorithmic} 
 \caption{Hamiltonian Construction: Dynamic Bit Masking}
 \label{alg:dynbit mask}
\end{algorithm}
\end{figure}

The dynamic bit masking algorithm is based on the property that any two bit strings that are at most doubly connected, differ by at most four orbitals. For a pair of bit strings that are doubly connected, there exists a set of four orbitals such that we can delete those orbitals and the two quadruply-deleted bit strings will become equivalent. Consider the example strings "111000011" and "111001100". By removing the first four orbitals (right to left ordering), both strings become "11100". Another way of doing this is simply to mask the bits (to zero), instead of deleting them. Either way, we will call the resulting bit string a 'reduced' bit string. Thus by checking all possible ways of removing four orbitals, we can determine if two bit strings are singly or doubly connected. 
To create an algorithm that is efficient and works for large number of bit strings, we have to expand this example in two essential ways. To compare a large list of reduced bit strings, after deleting the four orbitals in all bit strings, the reduced strings are sorted (which requires $\mathcal{O}(n\log(n))$ time), and then a final pass on the reduced strings is performed to look for blocks of adjacent reduced strings that are equal. The final pass can be $\mathcal{O}(n^{2})$ in the limit of dense matrices. For sparse Hamiltonians simulated in this work, this step is essentially linear in the number of determinants.

\begin{figure}
\begin{center}
\scalebox{1}{\includegraphics[width=1.0\columnwidth]{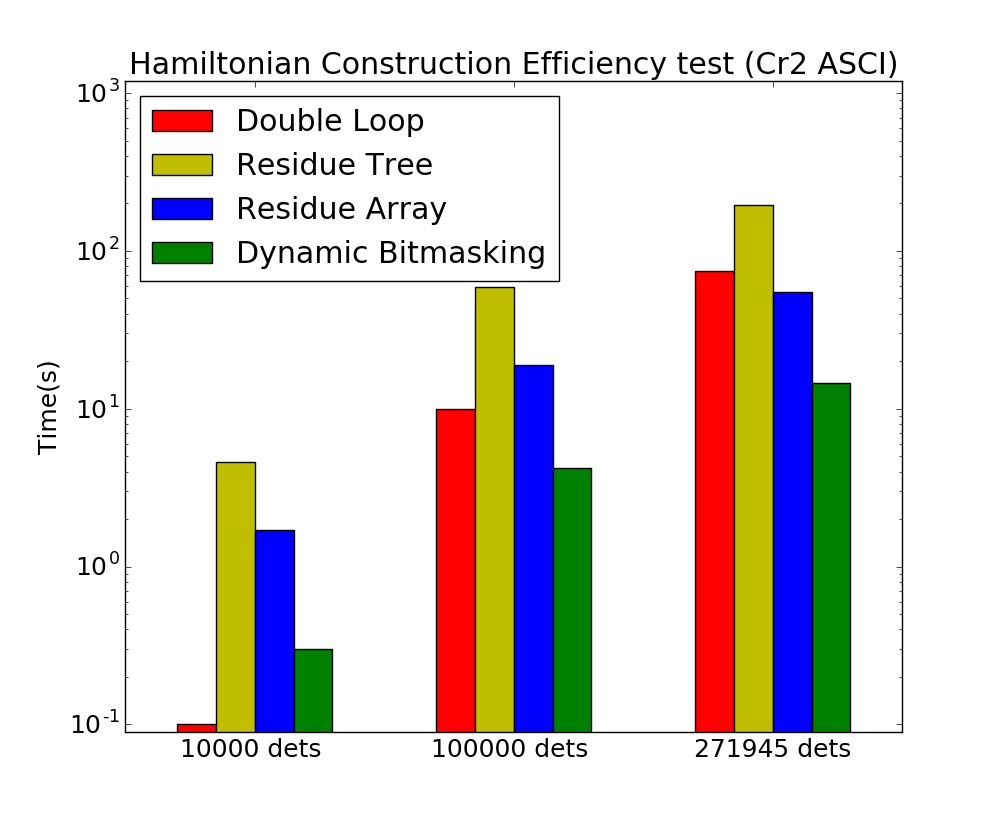}}
\end{center}
\caption{A comparison of different techniques to determine the non-zero matrix elements given a set of determinants from a Cr$_{2}$ ASCI simulation (smaller timings are better). 
The data is compared in detail in Table \ref{tab:speed}. Both dynamic bit masking (Algorithm \ref{alg:dynbit mask}) and residue arrays (Algorithm \ref{alg:residuearrays}) are new algorithms presented in this work and are based on sorting. The double loop algorithm is plotted as a baseline comparison, and is not competitive for large number of determinants.} 
\label{fig:speedhamil}
\end{figure}

\begin{table*}
\centering
\begin{tabular}{c||c|c|c|c|c||c|c|c|c|c||c|c|c|c|c||}
\# dets&\multicolumn{5}{c||}{$10000$}&\multicolumn{5}{c||}{$100000$}&\multicolumn{5}{c||}{$270000$}\\
\hline
&NNZ& DL & RT & RA & DB & NNZ& DL & RT & RA & DB &NNZ&DL & RT & RA &DB \\
\hline
Cr$_{2}$ ASCI & 412336& 0.1 & 4.6 & 1.7 & 0.3 & 7238098& 10 & 59 & 19 & 4.2 & 24095545& 75 & 196 & 55 & 14.5 \\
\hline
Cr$_{2}$ CISD & 5440608& 0.1 & 5.5 & 3.1 & 1.4 & 105791967& 10 & 84 & 43 & 29 & 469933565& 75 &261 & 144 & 147\\
\end{tabular}
\caption{Timing (in seconds) to construct the Hamiltonian with algorithms 2, 4, 5, 6, described in the text on two test cases: i) ASCI determinants for the Cr$_{2}$ SVP basis, and ii) a set of determinants that are generated from all singles and doubles of the Hartree-Fock reference for Cr$_{2}$ SVP. The ASCI test is also plotted in figure \ref{fig:speedhamil}. The abbreviations are as follows: DL (Double Loop, algorithm 2), RT (Residue Tree, algorithm 4), RA (Residue Array, algorithm 5), and DB (Dynamic Bit Masking, algorithm 6). NNZ stands for number of non-zero elements, and indicates the sparsity of the Hamiltonian. The number above each group of algorithms indicates the number of determinants used in the test. While still sparse, the singles and doubles test represents a much denser Hamiltonian than typically encountered in an ASCI simulation and it a worse case scenario for dynamic bit masking which takes advantage of Hamiltonian sparsity. Timings for a full Hamiltonian build, which includes the calculation of matrix elements, is presented in Table~\ref{tab:alltime}.}
\label{tab:speed}
\end{table*}%

The second key step is to reduce the number of quadruplets of orbitals to remove. In this current example, we search over all ways to remove four orbitals, of which there are $\binom{N_{orbs}}{4} $. Since this number can be quite large this approach would normally be a very slow way of constructing the Hamiltonian. To reduce the number of combinations of orbital removals, we use a coarse graining of the bit string by considering the bit string as non-overlapping substrings. If we break a pair of bit strings into some number $m$ of non-overlapping substrings 
(regardless of how this is done), at most four substrings can be different if the bit strings are singly or doubly connected. Hence all pairs of determinants, where the remaining $m-4$ substrings match, are candidates to have a non-zero matrix element between them. The trade off to this coarse graining is that there will be some false positives, which need to be checked for and removed. 

\begin{figure}
\begin{center}
\scalebox{1}{\includegraphics[width=1.0\columnwidth]{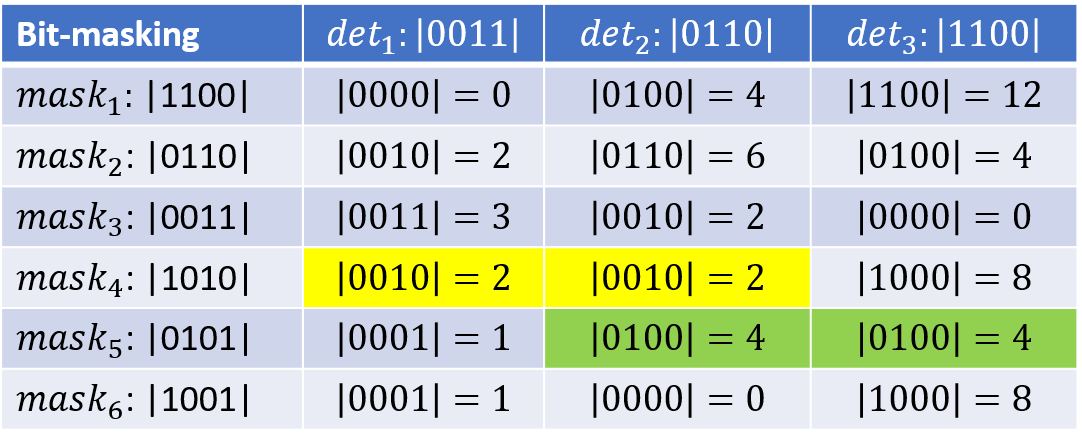}}
\end{center}
\caption{This table provides an example calculation for the logical $\rm{AND}$ of the $\binom{4}{2}$ bit masks that allow one to detect up to two differences between determinants. All bit masks are calculated for three sample determinants. Common entries along a row indicate that those determinants differ in no more than two places. These common entries are shaded in yellow and green to aid the eye.}
\label{fig:bit masktable}
\end{figure}

The way in which the bit strings are subdivided can be represented with bit masks. A simple but inefficient approach, as mentioned at the beginning of this section, is to have a bit mask for each orbital. However, it is easy to understand the approach in this limit. We provide a simple example for how bit masking can allow one to efficiently identify determinants that differ by up to two occupancies in figure~\ref{fig:bit masktable}.

Coarse-graining is most effective  when the the orbitals that are closest to being occupied 50\% of the time are put on different masks. Thus, one can pre-calculate orbital occupation over all the determinants in $\psi_k$. The "dynamic" part of this bit masking approach is named for this step where we use the wave function information to generate the bit masks. Once the bit masks have been selected, the algorithm proceeds as follows. For all determinants in the determinant list, compute the bitwise AND of the bit string representations of the determinants with the $m$ bit masks. Then, for all possible ways of selecting $m-4$ bit masks, create a unique value that is generated by the $m-4$ masked values. This can be done in any number of ways. A simple way is to create a combined integer by using shifting and adding operations to move all the masked values into the memory of a large integer. Sort this list of $N_\text{dets}$ combined integers to find repeated numbers, which correspond to non-zero elements of the Hamiltonian (checking for any false positives). For the tests we used in this work, we generally used 10 bit masks, which correspond to $\binom{10}{6} = 210$ different bit mask subsets. This corresponds to the number of sorting steps needed in the algorithm, which is one of the main costs. The more bit masks used, the fewer false positives. Timings for a full Hamiltonian build, which includes the calculation of matrix elements, are presented in section~\ref{sec:searchmain}. % and this is something that can be tuned for various Hamiltonians.

\begin{figure}
\centering
\includegraphics[width=1.0\columnwidth]{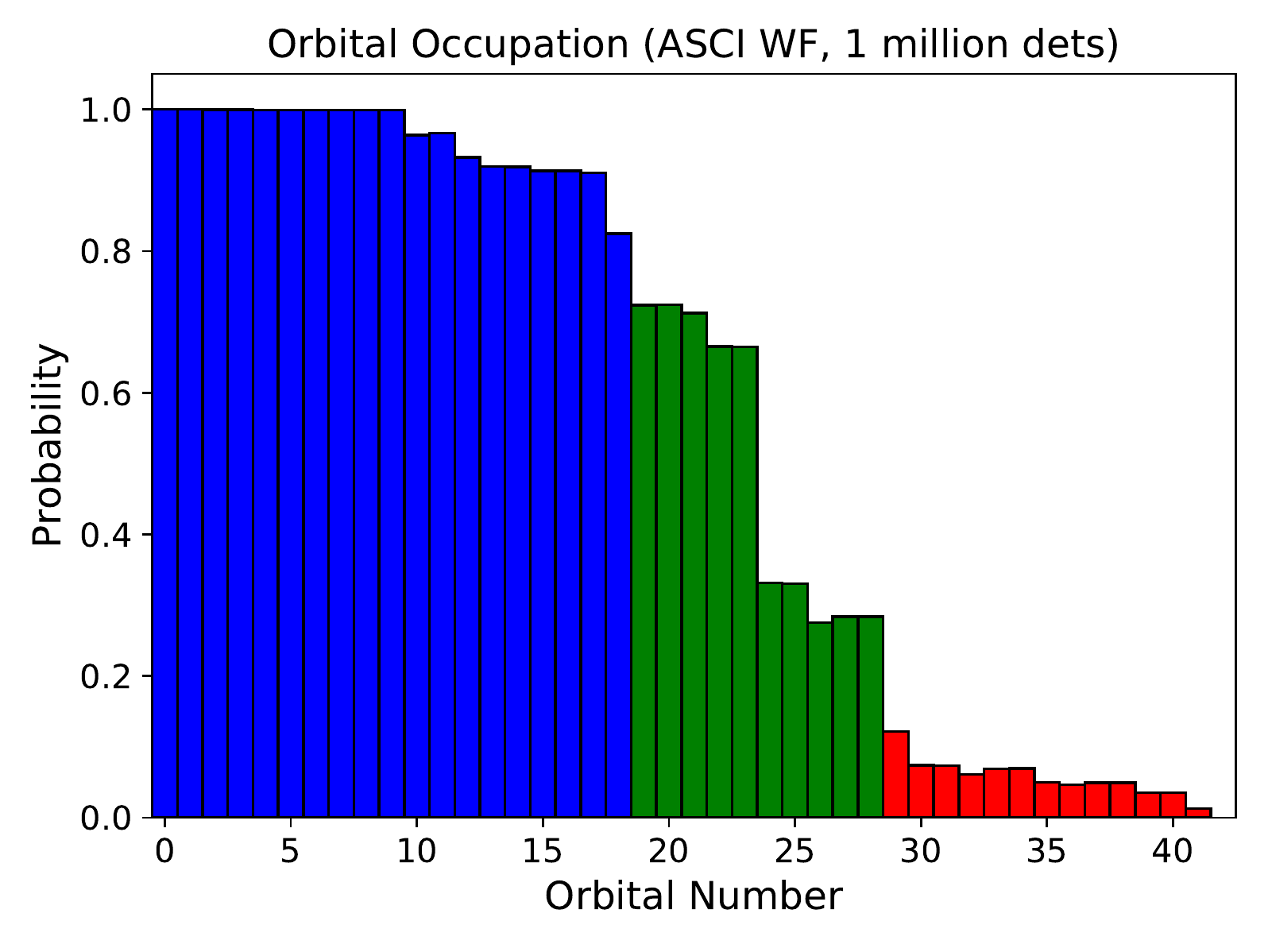}
\caption{Spin averaged orbital occupation plot (not weighted by determinant coefficient) for  Cr$_2$ in the SVP basis. There are 10 critical spatial orbitals (20 spin orbitals) that are near 50\% occupied, and are colored in green. The rest are mostly occupied (blue) or mostly empty (red). For our general Hamiltonian construction algorithms we fixed our calculations to 10 bit masks. Thus, for the Cr$_{2}$ SVP test, 2 critical orbitals are assigned to each bit mask. Conversely, none of the orbitals in the CISD wave function are near 50\% occupation, and therefore the orbitals can be assigned without much concern to the different bit masks.}
\label{fig:dynbit}
\end{figure}

\subsubsection{Hamiltonian Construction: Algorithm comparisons}
\label{sec:hcresults}

We summarize our main algorithmic results for the new diagonalization algorithms in figure~\ref{fig:speedhamil}. For these tests, we considered Hamiltonian construction for an all-electron Cr$_{2}$ SVP simulation as described in previous work~\cite{tubman2016-1}. 
%These tests are described in more detail in the relevant sections below (Section~\ref{sec:hamcons} and Section~\ref{sec:searchmain}). 
As noted above, residue arrays and dynamical bit masking are both based on sorting algorithms.  We have also tested many other approaches that do not use sorting algorithms, such as the residue tree algorithm~\cite{stampfub2005}. Search trees and hash tables, which are used in every current published implementation of selected CI that we are aware of, were found for the most part to be not competitive in any of our tests.  However, recent advances with hash tables and their related benchmarks need to be investigated further~\cite{hash_benchmark}.

\subsubsection{ Orbital rotations and calculating the 2RDM}

The dynamical bit masking technique is useful not only for calculating the Hamiltonian, but also for generating the 1-RDM and 2-RDM. In the case of a traditional full CI calculation, the single particle orbitals do not affect the final result. In a SCI simulation, which includes ASCI and HBCI, the single particle orbitals can influence the convergence of a simulation with regards to $N_{tdets}$. This suggests that single particle orbitals can be optimized and used to improve the energy convergence. In our initial ASCI implementation, we ran an iterative loop around our ASCI simulations, in which we performed orbital rotations using the 1-RDM at each step. At the end of a given ASCI run, the 1-RDM from the final wave function would be used to rotate to the natural orbital basis and the integrals would be recalculated. We would then use the new integrals as input for the next ASCI simulation. For C$_{2}$ and Cr$_{2}$, we found this procedure converged quite rapidly and needed fewer than four iterations of this in all the simulations presented here. 

In our current implementation we perform orbital rotations after each iteration during the wave function growth phase (see  figure~\ref{fig:flowchart}). After each diagonalization step in the growth phase, we rotate to the natural orbitals of the current wave function. We then immediately rediagonalize the Hamiltonian. This extra diagonalization step is important, since the expansion coefficients of the wave function are affected by the orbital rotation. After we finish the growth process, we fix the orbitals during the refinement steps.

\subsection{Search Algorithm}
\label{sec:searchmain}

\subsubsection{Search: new ASCI search}
\label{sec:searchtech}

The new ASCI search combines the pruning and ranking steps together such that the information generated in the pruning step is used to accelerate the ranking step. The new algorithm works as follows: generate all possible contributions of interest and store them in an array along with the generating determinant $i$. For an integral driven search, generate only those $\{D^{sd}\}$ such that $H_{ij} > \epsilon$. For a coefficient driven search, generate all $\{D^{sd}\}$ from the top $N_{cdets}$ determinants in $\{D_{}\}$. Either way, once the pruning is done, sort the array. With the sorted array, it is possible to calculate an approximate ranking of Eq.~\ref{eqn:rankeqn},
\begin{equation}
A_{i} = \frac{\sum^{'}_{j \ne i}H_{ij}C^k_{j}}{(H_{ii}-E_k)}.
\label{eqn:rankapprox}
\end{equation}
The prime on the sum indicates we are only including connections that fit our pruning criteria. %Note that in this approach we are combining the pruning and ranking steps together. We use the idea that we will find a connection multiple times to improve our calculation of the ranking equation, without slowing down the search algorithm too much.
For practical cases in which memory is limited, one can sort a partial array (after the maximum number of elements allowed in memory is reached), combine elements with the same bit string representation (that is, carry out a partial sum of Eq.~\ref{eqn:rankapprox}), and retain some percentage of the largest terms before continuing with the search. 

The ranking approximation developed for the HBCI were used in Ref.~\cite{holmes2016}
to reduce the cost of the search in the original ASCI algorithm. With the improved algorithms described here, this approximation is unnecessary (see Tables~\ref{tab:hbcomp} and ~\ref{tab:comparisons}). The accuracy of this new ASCI search is close  to that of the original ASCI search algorithm. In the HBCI ranking approximation, Eq.~\ref{eqn:rankeqn} was replaced with max($|C_{i}H_{ij}|$), which is too approximate and yields less accurate results. 

To further improve the computational efficiency of the new ASCI search approach, we speed up the calculation of the diagonal matrix elements, which are generally expensive to calculate and slow down the algorithm significantly (see Section~\ref{sec:fdiag}). The diagonal matrix elements are required for the denominator of Eq.~\ref{eqn:rankeqn}. The algorithm for calculating fast diagonal matrix elements is described in algorithm~\ref{alg:diag}. The new ASCI search algorithm (summarized in algorithm \ref{alg:ASCIsearchbetter}), is recommended for all compatible selected CI algorithms. In Table ~\ref{tab:hbcomp} and ~\ref{tab:comparisons}, we present a small number of comparisons of ASCI search to previously published HBCI and CIPSI results. In Table~\ref{tab:alltime}, we demonstrate that the new search algorithm is 
%almost never 
not a bottleneck in terms of computational time and can generally
be performed instead of less accurate searches.
See also Table~\ref{tab:speedsort} for generic timings of sorting integers. 

\begin{table}[]
\centering
\begin{tabular}{|l|l|l|l|l|l|}
\hline
\mc{2}{}& \mc{3}{Time (secs)}  &   \\\hline
System & Basis & H(Build) & H(Diag) & Search &NNZ(millions)  \\ \hline
OH (9e,44o)            & cc-pVTZ & 16   & 11      & 3   & 47 \\
CH (7e,44o)            & cc-pVTZ & 20   & 17      & 3   & 63 \\
Li$_{2}$(6e,60o)       & cc-pVTZ & 19   & 18      & 1   & 65 \\
Li$_{2}$(6e,110o)      & cc-pVQZ & 40   & 34      & 4   & 92 \\
F$_{2}$(18e,28o)       & cc-pVDZ & 6    & 2       & 1   & 14 \\
F$_{2}$(18e,60o)       & cc-pVTZ & 12   & 7       & 12  & 47 \\
N$_{2}$(14e,28o)       & cc-pVDZ & 10   & 10      & 1   & 25 \\
N$_{2}$(18e,60o)       & cc-pVTZ & 19   & 14      & 8   & 60 \\
C$_{2}$(12e,28o)       & cc-pVDZ & 8    & 6       & 1   & 21 \\
C$_{2}$(12e,60o)       & cc-pVTZ & 12   & 24      & 7   & 50 \\
C$_{2}$(12e,110o)      & cc-pVQZ & 26   & 32      & 48  & 68 \\ 
\hline
\end{tabular}
\caption{Timing of ASCI steps for various molecules. Timings presented for last step of growth algorithm in which 100,000 determinants are used. All timings are in seconds. NNZ is the number of non-zero matrix elements in the Hamiltonian. The column H(Build) is the timing for finding and calculating all non-zero matrix elements with dynamic bit masking. The column H(Diag) is the timing for matrix diagonalization with the Spectra package~\cite{spectracode,eigencode}. The search column is the timing for the ASCI search algorithm. These search timings do not represent the fastest set of parameters that make the search accurate and the timings would change slightly between a coefficient driven search versus an integral driven search. However it is evident that the timing for ASCI search is of the same order of magnitude as the other steps in ASCI and is not the bottleneck of the simulation. There is no reason to avoid using an accurate search algorithm such as the ASCI search. }
\label{tab:alltime}
\end{table}

\begin{figure}
\begin{algorithm}[H]
 \begin{algorithmic}[1]
\State Input $\psi_k$ and an array \{$V$\} that will hold information in pairs of (determinant bit string, ranking value ($RV$))
\State Order the determinants of $\psi_k$ by coefficient magnitude
%\State Order the two electron integrals by magnitude. For double connections these are equivalent to the $|H_{ij}|$ matrix elements.
\State Generate a set $D^{sd}$ from $\psi_k$ as follows: 
\begin{itemize}
\item Generate all single excitations $j$ in \{$D^{sd}$\} and calculate $C_{i}H_{ij}$
\item Generate double connections $j$ from \{$D^{sd}$\} with either a determinant driven or coefficient driven approach and calculate $c_{i}H_{ij}$
\item Add $j$ to an array \{$V$\} together with $RV$ =$c_{i}H_{ij}$
\item Sort the array $\{V\}$ by bit string. For all repeated elements $j$, sum up $RV$ to calculate the numerator of Eq.~\ref{eqn:rankeqn}
\item In a situation when memory is limited, perform a partial sort (by $RV$) and retain the top elements in array \{$V$\}, erase the rest of the array elements and continue with the search
\end{itemize}
\State Finish calculating Eq.~\ref{eqn:rankeqn} and gather all the top contributors for use in the next iteration
 \end{algorithmic} 
 \caption{ASCI Search}
 \label{alg:ASCIsearchbetter}
\end{algorithm}
\end{figure}

\subsection{Other algorithmic improvements}
\label{sec:misc}

In this work thus far, we have presented new algorithms that are efficient for performing the search and Hamiltonian construction. Most implementations of selected CI suffer significant performance penalties as a result of the growth of the data structures involved. In particular, the algorithms described in this work generally require the movement and manipulation of bit strings, which grow in size as more orbitals are included in a simulation. By studying and developing the sorting approaches contained in this work, we have been able to understand how most selected CI methods can benefit from cache efficient techniques. Additionally there are many helper algorithms that can be used in combination with the sorting algorithms described above to improve the over all efficiency.

\subsubsection{Fast Diagonal Matrix Elements}
\label{sec:fdiag}

During the calculation of the denominator of Eq.~\ref{eqn:rankeqn}, the diagonal matrix element of the connection being considered (H$_{ii}$) is required. Calculating the diagonal matrix elements is a relatively expensive step because these involve sums over the numbers of both electrons and pairs of electrons. However, because the determinant whose diagonal matrix element is being sought (H$_{ii}$) is always a single/double excitation away from a reference determinant, H$_{ref}$, for which the diagonal element is already known, the new matrix element H$_{ii}$  can be calculated quickly. This is to say that H$_{ref}$ - H$_{ii}$ only involves a small subset of terms, and can be calculated much faster than H$_{ii}$ 
from scratch. Algorithm \ref{alg:diag} describes this protocol.
Calculating the array of partial contributions described in algorithm \ref{alg:diag} only has to be done once per reference determinant.   This overhead  turns  out to be neglibile, as in either the PT2 or search algorithms, many connections from a reference determinant are considered all at once.  Thus the initial overhead is small compared to the number of diagonal matrix elements that need to be calculated for a given reference determinant.  

\begin{figure}
\begin{algorithm}[H]
 \begin{algorithmic}[1]
\State (Precalculation step) Input Determinant $D_{i}$, the diagonal matrix element $H_{ii}$, and the one-electron integrals $h_{ii}$
\State (Precalculation step) Calculate the partial contribution: $p(i) = \sum^{occ}_{j} \langle ij||ij\rangle$
\State Input $D_k$ (connected to $D_i$) with set of orbitals excited into ($A$) and the set of orbitals excited out of ($R$)
\State $E_{rem} = \sum_{i\in R}h_{ii}+p(i)$, $E_{add} = \sum_{i\in A}h_{ii}+p(i)$
\State $H_{kk}=H_{ii}-E_{rem}+E_{add}-\sum_{i\in R, j\in A}\langle ii||jj\rangle$
\If{$A, R$ have two elements from the same spin space}
$H_{kk} = H_{kk} + \langle R_{1}R_{1}||R_{2}R_{2}\rangle-\langle A_{1}A_{1}||A_{2}A_{2}\rangle-\langle R_{1}R_{1}||A_{2}A_{2}\rangle-\langle R_{2}R_{2}||A_{1}A_{1}\rangle$
\ElsIf{$A, R$ have two elements from different spin spaces}
$H_{kk} = H_{kk} + \langle R_{\alpha}R_{\alpha}|R_{\beta}R_{\beta}\rangle+\langle A_{\alpha}A_{\alpha}|A_{\beta}A_{\beta}\rangle-\langle A_{\alpha}A_{\alpha}|R_{\beta}R_{\beta}\rangle-\langle R_{\alpha}R_{\alpha}|A_{\beta}A_{\beta}\rangle$
\EndIf
 \end{algorithmic} 
 \caption{Fast Diagonal Matix elements}
 \label{alg:diag}
\end{algorithm}
\end{figure}

\subsubsection{Bit string representation}
Larger basis sets become more costly for ASCI for several reasons. One of the biggest costs comes from the cost associated with manipulating the larger bit strings associated with the larger basis set. Some of this cost can be mitigated by using more compact bit string representations. We first review the standard bit string representation before introducing selected CI adapted representations. 

Standard bit string representations use one bit for each spin orbital. Often these bits will be divided into a spin up string $\alpha$ and a spin down string $\beta$. When orbital $i$ is occupied by an $\alpha$ or $\beta$ electron, its bit will be set to 1 in either the alpha or beta string, respectively. For simplicity and alignment purposes within modern computing architectures, the number of bits is rounded up to the nearest power of 2. As an example the Cr$_{2}$ SVP basis set has 42 orbitals~\cite{yanai2009}. Thus a 64-bit integer would be used for the alpha and beta bit strings, combining to form a 128-bit string to represent the determinant.

\label{sec:ibit}

Comparisons and manipulations of larger integers take more time because larger blocks of memory must be compared and copied. Relatedly, parallel algorithms in which communication is the bottleneck also benefit from using smaller datatypes. Moreover, because large integers take up more space, fewer of them may be stored in the small cache available on the CPU. Additionally, most modern CPUs only implement 64-bit integer operations in hardware, so there is additional overhead required to perform bit manipulations in the multi-precision libraries that are implemented for handling integer types with greater than 64 bits. While the effect of the size of the representation of bit strings has not been explored in detail previously in the selected CI literature, this issue will occur in other selected CI approaches.

\begin{table}
\centering
\begin{tabular}{|c|c|c|}
\hline
Representation & Size & Description \\
\hline
Standard & $2(N_{orbs})$ & Regular\\
Electron & $N_{elec}*\log(N_{orbs})$&Electron occs\\
Difference & $N_{diff}\left(\log(N_{occ})+\log(N_{virt})\right)$ & HF diff\\
Hash & 64 (32 or 128 possible) & Use a 64 bit hash \\
\hline
\end{tabular}
\caption{A list of different bit string representations we considered in this work. See text for details.}
\label{tab:bitreps}
\end{table}%

In this work, we consider several different representations as well as an approach using hash functions. The length of these alternative representations (see Table~\ref{tab:bitreps}) can be much shorter than the standard representation. In the standard representation, 2*($N_{orbs}$) bits are used to represent all possible bit strings with $N_{orbs}$ spatial orbitals, regardless of how many quantum particles there are in the simulation.

In the \textit{electron representation}, rather than specifying whether orbitals are occupied or not, the occupied orbital indices are listed. That is, if orbitals 1, 2, and 10 are occupied, this determinant may be represented  by concatenating 1, 2, and 10 in binary. The amount of space required for this is $N_{elec}\log{N_{orb}}$ because the space required to store the maximum orbital index in binary is $\log{N_{orb}}$ and this is required for each electron. We use the convention that $\alpha$ electrons are concatenated before $\beta$ electrons, and, moreover, that for each spin the electrons are labeled in order from smallest to largest. This thus provides a unique bit string for a given determinant. The number of $\alpha$ and $\beta$ electrons does not have to be specified in each bit string since this is fixed for a given problem. This representation is best for systems with a small number of electrons but a large number of orbitals.

Another alternative is the \textit{difference representation}. In this representation, the determinant is specified by a list of the orbitals excited from the Hartree-Fock determinant and the orbitals into which they are excited. That is, if the electrons in orbitals 1 and 3 have been excited to 8 and 9, then the excited determinant is stored by concatenating 1, 3, 8, 9, and 2 in binary, with the final 2 added to specify that the particular determinant has two excitations. This representation is most useful when a large majority of the determinants are only a few excitations away from the Hartree-Fock determinant. Since the space required to store the maximum orbital excited out of and in to are $\log{N_{elec}}$ and $\log{N_{virt}}$, respectively, the memory requirements for this representation is $N_{diff}(\log(N_{occ})+\log(N_{virt})$. The number of bits needed can be reduced by only working with groups of bit strings that have fixed excitation number at one time. In such a situation, no bits are needed to represent the number of beta 
excitations, since this would be set by the number of alpha excitations and the total excitation number.

\subsubsection{Generalized compression with hash functions}
\label{sec:hbit}
The alternate bit string representations described above will not always reduce to the size of the standard bit string representation.
A more generic approach can be considered through hash functions. Hash functions are a 'many to one' map that reduces large data sets into fixed length integers and are widely known for their use in cryptography.  However hash functions also have wide usage for non-cryptographic purposes~\cite{smhasher1,smhasher2}.    The compression aspect of a hash function as well as the general property that hash functions are fast to calculate, are what we aim to exploit for enhancing the efficiency of our sorting algorithms.
%Compression in the case of selected CI is both highly desirable and always feasible. 
To understand this, it is important to note that for large basis set calculations, we will often make use of bit strings larger than 64 bits in order to represent a determinant.  For example there are over 10$^{36}$ unique numbers that can be represented with a 128 bit integer.  Yet we will only ever consider a very sparse number of determinants in this space.  Based on the timings in this paper, we might expect to simulate systems on the order of 10$^{13}$ determinants for a large scale PT2 calculation.  Thus for purposes for sorting, the question arises whether it is possible to hash (compress) large bit strings, such as 128 and larger, to a smaller size, such as 64 bits, without generating too many collisions. Some collisions are unlikely to effect the overall result, and it is straightforward to quantify how any might be generated. 
We will present a full benchmarking of hashing for large basis set simulations in future work.

\subsubsection{Perturbation Theory}
One of the most exciting aspects of the ASCI method is the ability to converge the energy with post processing of the final ASCI wave function, allowing one to produce a wave function that is better than virtually all other approximate CI techniques. %There have been many previous selected CI studies using CIPSI, however, none of them utilized perturbation theory to account for the unselected determinants and obtain near exact results until the ASCI method~\cite{tubman2016-1}. 
One of the main reasons that one might expect perturbation theory to be effective is that selected CI techniques finds all of the most important determinants in the Hilbert space and therefore anything that remains is necessarily small. There are multiple different ways in which the perturbation theory can be applied, including the Epstein-Nesbet perturbation theory that was introduced in Eq.~\ref{eqn:en}.
 
Recent approaches have tested the use of Monte Carlo sampling of these equations ~\cite{garniron2017, sharma2017}, and have been performed on roughly the same system sizes we presented here.
Within the selected CI community, there have been only limited attempts to make fast and scalable deterministic perturbation theory algorithms, since the focus has been on stochastic approaches.   However, we find that a deterministic approach will in many cases be faster than a stochastic approach in achieving chemical accuracy.  Indeed, if high accuracy is desired, the cost to converge the stochastic error of sampling techniques will likely be higher than that of deterministic approaches.  It should also be noted that many of the algorithmic improvements discussed in this work will also improve the stochastic approaches.
%, which makes a definitive comparison well outside the scope of this current work.  
In future studies we shall present the details of the deterministic PT algorithm we use in this work and a detailed discussion of where it is more efficient than stochastic methods.

\section{Results}
\label{sec:results}

\begin{table*}[]
\footnotesize
\centering
\begin{tabular}{|c|r|r|r||r|r|r||r|c|}
\hline
System           & E(10$^{5}$) & E(3*10$^{5}$) & E(10$^{6}$) & E(10$^{5}$)+PT2 & E(3*10$^{5}$)+PT2 & E(10$^{6}$)+PT2 & E(CCSD(T))   & ASCI-CCSD(T)(mHa) \\ \hline
\rowcolor{TABLE} 
BeH              & -15.189251  &               &             & -15.189270      &                   &                 & -15.189067 & -0.202            \\ \hline
C$_2$H$_2$       & -77.111633  & -77.113877    & -77.114973  & -77.116264      & -77.116281        & -77.116286      & -77.114792 & -1.493            \\ \hline
C$_2$H$_4$       & -78.347897  & -78.352665    & -78.356526  & -78.360534      & -78.360736        & -78.360870      & -78.359679 & -1.19             \\ \hline
C$_2$H$_6$       & -79.562899  & -79.569328    & -79.575622  & -79.586971      & -79.587362        & -79.587837      & -79.587598 & -0.238            \\ \hline
\rowcolor{TABLE} 
CH               & -38.381756  &               &             & -38.381830      &                   &                 & -38.381247 & -0.583            \\ \hline
\rowcolor{TABLE} 
CH$_2$\_singlet  & -39.024713  &               &             & -39.024890       &                   &                & -39.023981 & -0.908            \\ \hline
\rowcolor{TABLE} 
CH$_2$\_triplet  & -39.043515   &               &             &  -39.043647     &                   &                & -39.043196 & -0.451           \\ \hline
\rowcolor{TABLE} 
CH$_3$           & -39.718247  &               &             & -39.718698      &                   &                 & -39.718161 & -0.536            \\ \hline
CH$_3$Cl         &-499.429423  & -499.435363   & -499.440527 & -499.448171     & -499.448369       & -499.448604     & -499.447848 & -0.756            \\ \hline
CH$_4$           & -40.387284  & -40.388797    & -40.389683  & -40.390483      & -40.390461        & -40.390449      & -40.389881 & -0.567            \\ \hline
Cl$_2$           &-919.261710  & -919.267269   & -919.271444 & -919.275248     & -919.275433       & -919.275637     & -919.274483 & -1.154            \\ \hline
ClF              &-559.193146  & -559.197394   & -559.200247 & -559.203185     & -559.203196       & -559.203258     & -559.201909 & -1.349            \\ \hline
ClO              &-534.574415  & -534.578708   & -534.581433 & -534.583682     & -534.583783       & -534.583873     & -534.582017 & -1.856            \\ \hline
CN               & -92.494095  & -92.495493    & -92.496013  & -92.497001      & -92.496990        &  -92.496993     & -92.492776 & -4.217            \\ \hline
CO               &-113.055788  & -113.058002   & -113.05905  & -113.060133     & -113.060112       & -113.060115     & -113.058554 & -1.561            \\ \hline
CO$_2$           &-188.124313  & -188.133577   & -188.142101 & -188.155553     & -188.155768       & -188.155990     & -188.154316 & -1.673            \\ \hline
CS               &-435.606137  & -435.610029   & -435.612284 & -435.614274     & -435.614332       & -435.614381     & -435.612596 & -1.784            \\ \hline
F$_2$            &-199.096831  & -199.099912   & -199.101763 & -199.103291     & -199.103314       & -199.103345     & -199.101481 & -1.863            \\ \hline
H$_2$CO          &-114.213577  & -114.217838   & -114.221088 & -114.224420      & -114.224415       & -114.224453     & -114.222990 & -1.462            \\ \hline
\rowcolor{TABLE} 
H$_2$O           & -76.243432  &               &             & -76.243908      &                   &                 & -76.243266 & -0.641            \\ \hline
H$_2$O$_2$       & -151.180192 & -151.185981   & -151.191415 & -151.199403     & -151.199388       & -151.199442     & -151.197870 & -1.571            \\ \hline
H$_2$S           & -398.871385 & -398.871798   & -398.871891 & -398.872360      & -398.872357       & -398.872357     & -398.871682 & -0.675            \\ \hline
H$_3$COH         & -115.403262 & -115.409733   & -115.415589 & -115.425482     & -115.425525       & -115.425695     & -115.424943 & -0.752            \\ \hline
H$_3$CSH         & -438.038163 & -438.045231   & -438.051720  & -438.063596     & -438.063899       & -438.064300     & -438.063831 & -0.469            \\ \hline
\rowcolor{TABLE} 
HCl              & -460.260248 &               &             & -460.26072      &                   &                 & -460.260217 & -0.502            \\ \hline
HCN              & -93.189422  & -93.192245    & -93.193804  & -93.195281      & -93.195268        & -93.195273      & -93.193547 & -1.726            \\ \hline
HCO              & -113.571362 & -113.575611   & -113.578885 & -113.582082     & -113.58199        & -113.581967     & -113.580181 & -1.785            \\ \hline
\rowcolor{TABLE} 
HF               & -100.229942 &               &             & -100.230383     &                   &                 & -100.229878 & -0.504            \\ \hline
HOCl             & -535.221695 & -535.227183   & -535.231958 & -535.237723     & -535.237791       & -535.237901     & -535.236561 & -1.34             \\ \hline
\rowcolor{TABLE} 
Li$_2$           & -14.901321  &               &             & -14.901337      &                   &                 & -14.901331 & -0.005            \\ \hline
\rowcolor{TABLE} 
LiF              & -107.157186 &               &             & -107.157874     &                   &                 & -107.157255 & -0.618            \\ \hline
\rowcolor{TABLE} 
LiH              & -8.014688   &               &             & -8.0147070      &                   &                 & -8.014708  & 0.002             \\ \hline
N$_2$            & -109.279470 & -109.280666   & -109.280941 & -109.281912     & -109.281927       & -109.281933     & -109.279982 & -1.95             \\ \hline
N$_2$H$_4$       & -111.547435 & -111.554763   & -111.561876 & -111.575558     & -111.575552       & -111.575733     & -111.575055 & -0.678            \\ \hline
\rowcolor{TABLE} 
Na$_2$           & -323.733949 &               &             & -323.733997     &                   &                 & -323.734047 & 0.051             \\ \hline
\rowcolor{TABLE} 
NaCl             & -621.595166 &               &             & -621.595793     &                   &                 & -621.595302 & -0.49             \\ \hline
\rowcolor{TABLE} 
NH               & -55.093412  &               &             & -55.093530       &                   &                 & -55.093131 & -0.399            \\ \hline
\rowcolor{TABLE} 
NH$_2$           & -55.735042  &               &             & -55.735304      &                   &                 & -55.734737 & -0.567            \\ \hline
NH$_3$           & -56.403846  & -56.404593    & -56.404853  & -56.405301      & -56.405298        & -56.405299      & -56.404668 & -0.631            \\ \hline
NO               & -129.597925 & -129.600605   & -129.601836 & -129.603099     & -129.60308        & -129.603079     & -129.60115  & -1.926            \\ \hline
O$_2$            & -149.986002 & -149.988413   & -149.989404 & -149.990669     & -149.990697       & -149.990712     & --149.988242 & -2.469            \\ \hline
\rowcolor{TABLE} 
OH               & -75.561403  &               &             & -75.561639      &                   &                 & -75.561190 & -0.449            \\ \hline
P$_2$            & -681.733926 & -681.736715   & -681.738451 & -681.740076     & -681.740139       & -681.740138     & -681.737699 & -2.439            \\ \hline
\rowcolor{TABLE} 
PH$_2$           & -342.015274 &               &             & -342.015912     &                   &                 & -342.015208 & -0.703            \\ \hline
PH$_3$           & -342.643041 & -342.644264   & -342.644881 & -342.645539     & -342.645537       & -342.645538     & -342.644777 & -0.76             \\ \hline
S$_2$            & -795.334164 & -795.338687   & -795.341495 & -795.344591     & -795.344788       & -795.344873     & -795.343001 & -1.872            \\ \hline
Si$_2$           & -577.937111   &-577.938035   & -577.938624 & -577.940646     & -577.940670
       & -577.940685
     & -577.938371 & -2.314            \\ \hline
Si$_2$H$_6$      & -581.596630 & -581.602802   & -581.609336 & -581.622553     & -581.623534       & -581.623600       & -581.625045 & 1.445             \\ \hline
\rowcolor{TABLE} 
SiH$_2$\_singlet & -290.143803 &               &             & -290.144185     &                   &                 & -290.143496 & -0.689            \\ \hline
\rowcolor{TABLE} 
SiH$_2$\_triplet & -290.101056 &               &             & -290.101377     &                   &                 & -290.100697 & -0.68             \\ \hline
SiH$_3$          & -290.754016 & -290.754683   & -290.755020 & -290.755380     & -290.755385       & -290.755393     & -290.754754 & -0.64             \\ \hline
SiH$_4$          & -291.396355 & -291.398117   & -291.399246 & -291.400364     & -291.400415       & -291.400435     & -291.399825 & -0.61             \\ \hline
SiO              & -364.086144 & -364.088978   & -364.090533 & -364.092085     & -364.092087       & -364.092108     & -364.090065 & -2.042            \\ \hline
SO               & -472.662216 & -472.666953   & -472.670542 & -472.673837     & -472.673857       & -472.673893     & -472.671759 & -2.134            \\ \hline
SO$_2$           & -547.683822 & -547.696538   & -547.707992 & -547.732183     & -547.732764       & -547.733404     & -547.731453 & -1.95             \\ \hline
\end{tabular}
\caption{Ground state energies for the G1 molecules in a cc-pVDZ basis. The ASCI results (columns 1 - 6) are labeled by the number of determinants, with both variational and perturbation results presented. The best perturbation results (column 6) are compared with CCSD(T) (column 7). Some of the systems have small Hilbert spaces and showed convergence within 1 mHa between the variational and PT2 results, with less than 10$^{5}$ determinants (green rows). All geometries are taken from the original G1 set~\cite{pople1989} except for CN and CH$_{2}$ triplet, for which we use the geometries from Ref.~\cite{feller1}.}
\label{tab:dz}
\end{table*}

\begin{table*}[]
\footnotesize
\centering
\begin{tabular}{|c|r|r|r||r|r|r||r|c|}
\hline
System           & E(10$^{5}$) & E(3*10$^{5}$) & E(10$^{6}$) & E(10$^{5}$)+PT2 & E(3*10$^{5}$)+PT2 & E(10$^{6}$)+PT2 & E(CCSD(T))  & ASCI-CCSD(T)(mHa) \\ \hline
\rowcolor{TABLE} 
BeH              & -15.202968  &               &             & -15.203059      &                   &                 & -15.202841  & -0.218            \\ \hline
C$_2$H$_2$       & -77.203740   & -77.209351    & -77.213467  & -77.219532      & -77.219470         & -77.219436      & -77.218046  & -1.39             \\ \hline
C$_2$H$_4$       & -78.441654  & -78.450402    & -78.457561  & -78.471004      & -78.471438        &                 & -78.470558  & -0.88             \\ \hline
C$_2$H$_6$       & -79.654252  & -79.669309    & -79.680275  & -79.706885      & -79.707123        &                 & -79.707843  & 0.72              \\ \hline
\rowcolor{TABLE} 
CH               & -38.422000  &               &             & -38.422323      &                   &                & -38.421572  &             -0.751\\ \hline
CH$_2$\_singlet  & -39.074412  & -39.074996    & -39.075219  & -39.075663      & -39.075638        & -39.075641      & -39.074555  & -1.086            \\ \hline
\rowcolor{TABLE} 
CH$_2$\_triplet  & -39.091649   &     &    & -39.092561      &          &      & -39.091962   &     -0.599        \\ \hline
CH$_3$           & -39.774767  & -39.775888    & -39.776526  & -39.777381      & -39.777335        & -39.777308      & -39.776688  & -0.62             \\ \hline
CH$_3$Cl         & -499.558732 & -499.574866   & -499.586523 & -499.613409     & -499.613956       &                 & -499.614776 & 0.82              \\ \hline
CH$_4$           & -40.445943  & -40.449959    & -40.452376  & -40.455999      & -40.455844        & -40.455727      & -40.455065  & -0.662            \\ \hline
Cl$_2$           & -919.443813 & -919.456931   & -919.467067 & -919.489489     & -919.490563       &                 & -919.491057 & 0.494             \\ \hline
ClF              & -559.385351 & -559.397209   & -559.406478 & -559.424939     & -559.425259       & -559.425479     & -559.424306 & -1.173            \\ \hline
ClO              & -534.743221 & -534.75429    & -534.763177 & -534.779342     & -534.779876       & -534.780297     & -534.778399 & -1.898            \\ \hline
CN               & -92.579322  & -92.585386    & -92.589739  & -92.595444      & -92.595278        & -92.595193     & -92.5907969 & -4.396            \\ \hline
CO               & -113.164548 & -113.170564   & -113.174971 & -113.181556     & -113.18131        & -113.181203     & -113.179827 & -1.376            \\ \hline
CO$_2$           & -188.30436  & -188.320873   & -188.334374 & -188.366900       & -188.367270        &                 & -188.367412 & 0.142             \\ \hline
CS               & -435.711205 & -435.720092   & -435.727456 & -435.739697     & -435.740067       & -435.740301     & -435.738264 & -2.037            \\ \hline
F$_2$            & -199.297893 & -199.305614   & -199.311568 & -199.322174     & -199.322044       & -199.322013     & -199.320490  & -1.523            \\ \hline
H$_2$CO          & -114.333293 & -114.342815   & -114.349969 & -114.363196     & -114.363265       & -114.363298     & -114.362161 & -1.137            \\ \hline
H$_2$O           & -76.342531  & -76.344107    & -76.344966  & -76.346114      & -76.346046        & -76.346017      & -76.345555  & -0.462            \\ \hline
H$_2$O$_2$       & -151.328267 & -151.346149   & -151.358343 & -151.38593      & -151.385453       &                 & -151.383997 & -1.456            \\ \hline
H$_2$S           & -398.963339 & -398.966601   & -398.968840  & -398.971854     & -398.971758       & -398.971672     & -398.970663 & -1.009            \\ \hline
H$_3$COH         & -115.518981 & -115.538789   & -115.551947 & -115.581707     & -115.581297       &                 & -115.580557 & -0.74             \\ \hline
H$_3$CSH         & -438.147677 & -438.169313   & -438.184286 & -438.219185     & -438.219443       &                 & -438.220721 & 1.278             \\ \hline
HCl              & -460.365313 & -460.367443   & -460.368720  & -460.370542     & -460.370467       & -460.370428     & -460.369640  & -0.788            \\ \hline
HCN              & -93.283038  & -93.290436    & -93.295931  & -93.304741      & -93.304506        & -93.304428      & -93.302782  & -1.646            \\ \hline
HCO              & -113.679998 & -113.690685   & -113.698435 & -113.712574     & -113.712376       & -113.712248     & -113.710567 & -1.681            \\ \hline
\rowcolor{TABLE} 
HF               & -100.349364 &               &             & -100.351308     &                   &                 & -100.351011 & -0.297            \\ \hline
HOCl             & -535.38483  & -535.401621   & -535.413545 & -535.439971     & -535.440168       &                 & -535.439573 & -0.595            \\ \hline
\rowcolor{TABLE} 
Li$_2$           & -14.930785  &               &             & -14.930786      &                   &                 & -14.930734  & -0.05             \\ \hline
LiF              & -107.288594 & -107.289889   & -107.290477 & -107.291433     & -107.291471       & -107.291499     & -107.291267 & -0.232            \\ \hline
\rowcolor{TABLE} 
LiH              & -8.036373   &               &             & -8.036477       &                   &                 & -8.036468   & -0.009            \\ \hline
N$_2$            & -109.386980  & -109.391765   & -109.395149 & -109.399681     & -109.399578       & -109.399513     & -109.397769 & -1.744            \\ \hline
N$_2$H$_4$       & -111.653829 & -111.677219   & -111.693657 & -111.730173     & -111.728936       &                 & -111.727644 & -1.292            \\ \hline
\rowcolor{TABLE} 
Na$_2$           & -323.768264 &               &             & -323.769296     &                   &                 & -323.769243 & -0.053            \\ \hline
NaCl             & -621.715212 & -621.718356   & -621.720573 & -621.723618     & -621.723678       & -621.723662     & -621.723066 & -0.596            \\ \hline
\rowcolor{TABLE} 
NH               & -55.152475  &               &             & -55.153059      &                   &                 & -55.152497  & -0.562            \\ \hline
NH$_2$           & -55.805235  & -55.806328    & -55.806869  & -55.807663      & -55.807618        & -55.807603      & -55.806951  & -0.652            \\ \hline
NH$_3$           & -56.482279  & -56.484996    & -56.486656  & -56.489049      & -56.488902        & -56.488806      & -56.488205  & -0.601            \\ \hline
NO               & -129.723293 & -129.730435   & -129.73569  & -129.743992     & -129.743756       & -129.743632     & -129.741868 & -1.764            \\ \hline
O$_2$            & -150.128809  & -150.137355    & -150.144920  &-150.154261      & -150.154042       & -150.154011     & -150.151869 & -2.142             \\ \hline
\rowcolor{TABLE} 
OH               & -75.649012  &     &   & -75.650479      &         &        & -75.649945  & -0.521             \\ \hline
P$_2$            & -681.858265 & -681.867251   & -681.875496 & -681.891685     & -681.892673       & -681.893154     & -681.890509 & -2.645            \\ \hline
PH$_2$           & -342.093796 & -342.096363   & -342.098130  & -342.100204     & -342.100126       & -342.100058     & -342.099026 & -1.032            \\ \hline
PH$_3$           & -342.724455 & -342.729358   & -342.732548 & -342.737893     & -342.737852       & -342.737824     & -342.736695 & -1.129            \\ \hline
S$_2$            & -795.479446 & -795.492316   & -795.501795 & -795.523482     & -795.524441       &                 & -795.52348  & -0.961            \\ \hline
Si$_2$ & -578.064458         & -578.073053 & -578.078038   & -578.091580 &    -578.092361          &    -578.0927272             & -578.090310 & -2.417            \\ \hline
Si$_2$H$_6$      & -581.742988 & -581.757872   &             & -581.808668     & -581.810282       &                 & -581.816405 & 6.123             \\ \hline
SiH$_2$\_singlet & -290.224640  & -290.226524   & -290.227658 & -290.228906     & -290.228859       & -290.228837     & -290.227976 & -0.861            \\ \hline
SiH$_2$\_triplet & -290.179533 & -290.181296   & -290.182352 & -290.183480      & -290.183430        & -290.183409     & -290.182552 & -0.857            \\ \hline
SiH$_3$          & -290.836906 & -290.840366   & -290.842778 & -290.845976     & -290.845983       & -290.845917     & -290.845112 & -0.805            \\ \hline
SiH$_4$          & -291.482012 & -291.487852   & -291.492150  & -291.498424     & -291.498589       & -291.498717     & -291.498002 & -0.715            \\ \hline
SiO              & -364.239919 & -364.247764   & -364.254334 & -364.265039     & -364.264860        & -364.264922     & -364.263351 & -1.571            \\ \hline
SO               & -472.815828 & -472.828155   & -472.837710  & -472.857033     & -472.857111       & -472.857246     & -472.855188 & -2.058            \\ \hline
SO$_2$           & -547.901866 & -547.936060   & -547.958730  & -548.026183     & -548.026880        &                 & -548.030305 & 3.425             \\ \hline
\end{tabular}
\caption{Ground state energies of the G1 molecules in a cc-pVTZ basis. The ASCI results are labeled by the number of determinants (columns 1-6), with both variational and perturbation results presented. The best perturbation results (column 6) are compared with CCSD(T). With
the exception of Si$_2$H$_6$, we were able to run PT2 for all systems with at least 3*10$^{5}$ determinants within the time limit of our computational resources (see text for details of these). }
\label{tab:tz}
\end{table*}

\begin{table}[htb!]

\centering
\begin{tabular}{|c|r|r|}
\hline
Atoms&E(cc-pVDZ)&E(cc-pVTZ) \\
\hline
B   &   -24.59062  &  -24.60580  \\
Be  &   -14.61740  &  -14.62379 \\
C   &   -37.76190  &  -37.79003  \\
Cl  &  -459.60432  & -459.70385  \\
F   &   -99.52947  &  -99.63240  \\
H   &    -0.49927  &   -0.49980  \\
Li  &    -7.43263  &   -7.44606   \\
N   &   -54.48011  &  -54.52523   \\
Na  &  -161.85418  & -161.86990  \\
O   &   -74.91171  &  -74.98526   \\
P   &  -340.79727  & -340.86128    \\
S   &  -397.60643  & -397.68672   \\
Si  &  -288.92066  & -288.98829   \\
\hline
\end{tabular}
\caption{Atomic energies for cc-pVDZ and cc-pVTZ atoms in the G1 set.  All energies are in units of Ha.  All atomic energy calculations are ASCI+PT2 energies.
These atomic energies are very easy to simulate with ASCI and all of the presented results are more than 0.1 mHa accurate. 
}
\label{tab:atomene}
\end{table}

\begin{table*}[]
\centering
\begin{tabular}{|l|l|l|l|l||l|l|l|l|l|}
\hline
System          & \begin{tabular}{@{}c@{}}ASCI D$_{e}$\\  cc-pVDZ\end{tabular} 
  & \begin{tabular}{@{}c@{}}CCSDTQ D$_{e}$\\  cc-pVDZ~\cite{feller1}\end{tabular} &  \begin{tabular}{@{}c@{}}ASCI D$_{e}$\\  cc-pVTZ\end{tabular}  &  \begin{tabular}{@{}c@{}}CCSD(T) D$_{e}$\\  CBS FC~\cite{feller1,feller2}\end{tabular} & System          & \begin{tabular}{@{}c@{}}ASCI D$_{e}$\\  cc-pVDZ\end{tabular} 
  & \begin{tabular}{@{}c@{}}CCSDTQ D$_{e}$\\  cc-pVDZ~\cite{feller1}\end{tabular} &  \begin{tabular}{@{}c@{}}ASCI D$_{e}$\\  cc-pVTZ\end{tabular}  &  \begin{tabular}{@{}c@{}}CCSD(T) D$_{e}$\\  CBS FC~\cite{feller1,feller2}\end{tabular}\\ \hline
BeH             & 45.546          &                    & 49.858          & 50.11                  & HOCl             & 139.668         &                   & 157.654         &   165.5                    \\ \hline
C$_2$H$_2$      & 372.692         & 371.71             & 401.442         & 402.78                 & Li$_2$           & 22.635          &                  & 24.245          & 24.49                  \\ \hline
C$_2$H$_4$      & 527.077         &                    & 560.148         & 561.51                 & LiF              & 122.846         &                   & 133.676         &      137.7                 \\ \hline
C$_2$H$_6$      & 670.406         &                    & 707.948         & 710.5                  & LiH              & 51.955          &                   & 56.852          & 58.11                  \\ \hline
CH              & 75.707          &                    & 83.128          & 83.89                  & N$_2$            & 201.877         & 201.44             & 219.031         & 227.14                 \\ \hline
CH$_2$\_singlet & 165.932         & 165.83             & 179.457         & 180.68                 & N$_2$H$_4$       & 388.051         &                   & 426.224         & 436.87                 \\ \hline
CH$_2$\_triplet & 177.702         & 177.6              & 190.074         & 189.85                 & Na$_2$           & 16.081          &                   & 18.504          &   16.7                     \\ \hline
CH$_3$          & 288.002         & 287.62             & 306.124         & 306.71                 & NaCl             & 86.144          &                   & 94.063          &        99.3               \\ \hline
CH$_3$Cl        & 366.802         &                    & 389.450          & 394.83                 & NH               & 71.625          & 71.61              & 80.331          & 82.85                  \\ \hline
CH$_4$          & 396.229         &                    & 418.203         & 419.14                 & NH$_2$           & 161.042         &                   & 177.429         & 182                    \\ \hline
Cl$_2$          & 42.029          & 42.13              & 51.987          & 59.87                  & NH$_3$           & 268.168         &                   & 291.255         & 297.2                  \\ \hline
ClF             & 43.584          &                    & 55.984          & 62.61                  & NO               & 132.562         & 132.8              & 146.296         & 151.77                 \\ \hline
ClO             & 42.561          &                    & 57.215          & 64.56                  & O$_2$            & 104.967         & 105.19             & 115.166         & 119.9                  \\ \hline
CN              & 160.001        & 160.3              & 175.653         & 179.2                  & OH               & 94.530           & 94.4               & 101.028         & 107.06           \\ \hline
CO              & 242.529         & 242.04             & 254.708         & 258.59                 & P$_2$            & 91.361          & 90.43              & 107.038         & 115.85                 \\ \hline
CO$_2$          & 358.089         &                    & 380.715         & 388.12                 & PH$_2$           & 138.104         &                    & 150.068         &      154.0                  \\ \hline
CS              & 154.395         & 154.1              & 165.372         & 170.93                 & PH$_3$           & 219.898         &                    & 236.637         & 241.65                 \\ \hline
F$_2$           & 27.860           & 28.19              & 35.896          & 38.43                  & S$_2$            & 82.836          & 82.43              & 94.750           & 103.52                 \\ \hline
H$_2$CO         & 346.558         & 345.79             & 369.213         & 373.15                 & Si$_2$           & 61.811          & 61.94              & 70.958          & 75.95                  \\ \hline
H$_2$O          & 209.359         & 209.16             & 226.616         & 232.67                 & Si$_2$H$_6$      & 493.601         &                    & 523.189         &  535.0                      \\ \hline
H$_2$O$_2$      & 236.853         &                    & 260.611         & 268.32                 & SiH$_2$\_singlet & 141.169         & 140.64             & 151.183         &      153.9                   \\ \hline
H$_2$S          & 167.776         &                    & 179.047         & 183.51                 & SiH$_2$\_triplet & 15.510           &                    & 122.676         &        133.5                \\ \hline
H$_3$COH        & 473.745         &                    & 506.248         &  510.9                      & SiH$_3$          & 211.405         &                    & 224.771         &    228.7                    \\ \hline
H$_3$CSH        & 438.535         &                    & 466.516         &   472.3                     & SiH$_4$          & 302.872         &                    & 320.774         &    324.3                    \\ \hline
HCl             & 98.588          & 98.29              & 104.643         & 107.39                 & SiO              & 162.981         &                    & 182.836         & 191.77                 \\ \hline
HCN             & 284.877         & 283.38             & 307.070          & 311.37                 & SO               & 97.730           & 97.22              & 116.253         & 125.73                 \\ \hline
HCO             & 256.693         &                    & 274.308         &    277.1                    & SO$_2$           & 190.471         &                    & 231.949         & 259.14                 \\ \hline
HF              & 126.525         & 126.52             & 137.468         & 141.59                 &                  &                  &                    &                  &                        \\ \hline
\end{tabular}
\caption{Atomization energies from ASCI, compared to benchmark results. The coupled cluster results are from references~\cite{feller1,feller2}. The geometries used in these references are slightly different than the geometries used in this work. However, for cc-pVDZ, the ASCI results compared to the  CCSDTQ show strong agreement, less than 1 kcal/mol accurate across all provided results. The ASCI (cc-pVTZ,all electron) results compared to the CCSD(T)(CBS,Frozen Core) are presented to make a qualitative comparison of the convergence of ASCI/cc-pVTZ to the CBS limit. }
\label{tab:atom}
\end{table*}

\begin{table}[]
\centering
\begin{tabular}{|l|l|l|}
\hline
System            &  E(10$^{6}$)+PT2 (cc-pVDZ) & E(3*10$^{5}$)+PT2 (cc-pVTZ)
\\ \hline
C$_2$H$_4$        &  -78.35920      &   -78.47061    \\ \hline
C$_2$H$_6$        &  -79.58738      & -79.70721      \\ \hline
CH$_3$Cl          &  -499.44827      & -499.61401      \\ \hline
CH$_4$            &  -40.39016       & -40.45587      \\ \hline
Cl$_2$            &  -919.27460       & -919.48997      \\ \hline
ClF               &  -559.20164       & -559.42533      \\ \hline
ClO               &  -534.58136      &  -534.77956     \\ \hline
CO                &  -113.06008     & -113.18187      \\ \hline
CO$_2$            &  -188.15522      & -188.36808       \\ \hline
CS                &  -435.61384     & -435.73994       \\ \hline
F$_2$             &  -199.10299     &-199.32210      \\ \hline
H$_2$CO           &  -114.22396      & -114.36354      \\ \hline
H$_2$O$_2$        &  -151.19915      & -151.38581     \\ \hline
H$_2$S            &  -398.87226      & -398.97178      \\ \hline
HCN               &  -93.19409       & -93.30518      \\ \hline
HCO               &  -113.58083      & -113.71129      \\ \hline
HOCl              &  -535.23687      & -535.44029      \\ \hline
N$_2$             &  -109.28088      & -109.40151      \\ \hline
NH$_3$            &  -56.40492       &  -56.48895      \\ \hline
NO                &  -129.60369      &  -129.74398     \\ \hline
P$_2$             &  -681.73884      & -681.89253      \\ \hline
PH$_3$            &  -342.64550     & -342.73796      \\ \hline
Si$_2$            &  -577.94012     & -578.09017      \\ \hline
SiO               &  -364.09010      & -364.26490      \\ \hline
SO                &  -472.67119      &  -472.85738     \\ \hline
SO$_2$            &  -547.72710      &  -548.02811     \\ \hline
\end{tabular}
\caption{Ground state energies calculated with ASCI and PT2 corrections for selected molecules using geometries from Feller \textit{et. al.}~\cite{feller1}.  Energies are presented in units of Ha.  Most of the energies shown here are within 2 mHa of the energies calculated with the geometries from the original G1 set that are shown in Tables~\ref{tab:dz} and \ref{tab:tz}.}
\label{tab:feller}
\end{table}

We present several benchmark studies below in order to provide an overview of what can be done with selected CI today. The algorithms described above have been implemented in our own code~\cite{tubman2016-1} and in QChem 5.1~\cite{shao_advances_2015}. With the exception of a few instances, the results in this section were generated on a single core with an upper limit of 50 CPU (single core) hours. Many of these calculations were made over the course of developing the methods in these papers.  

\subsection{G1 dataset}
%\textit{A:  G2 set}
%To demonstrate a benchmark set of simulations for selected CI, we 
Tables~\ref{tab:dz} and~\ref{tab:tz} present ASCI results on the G1 test set of 55 molecules (all electron),  a benchmark set of molecules that has been extensively studied with many different methods\cite{pople1989,feller1,feller2,grossman2002,ma2005,mardirossian2017thirty,hait2018accurate}. Here we present ASCI benchmark data with both cc-pVDZ and cc-pVTZ basis sets. The molecular geometries were taken from the original G1 set~\cite{pople1989,grossman2002,ma2005}, except for CN and CH$_{2}$ triplet where we use the geometries from Feller \textit{et. al.}~\cite{feller1}.  ASCI simulations with different numbers of determinants were performed to demonstrate the convergence with respect to this parameter. 

Tables~\ref{tab:dz} and~\ref{tab:tz} show that nearly all of the molecules are converged to within chemical accuracy for the cc-pVDZ simulations and many are also converged in the cc-pVTZ basis set. These results suggest that cc-pVQZ convergence will be possible for all the G1 molecules in the near future, using only modest computational resources.

\begin{figure}
\centering{\includegraphics[width=1.0\columnwidth]{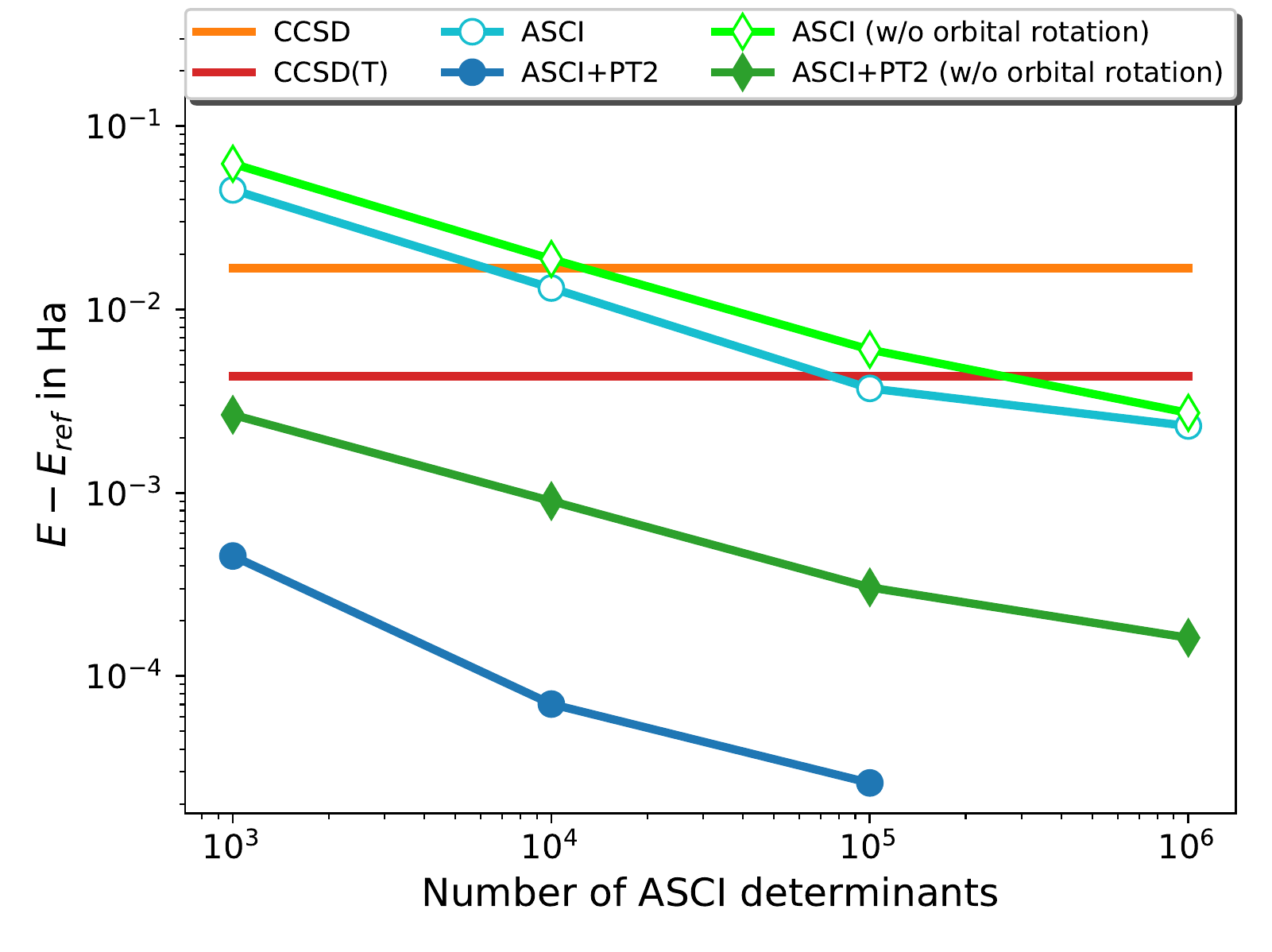}}
\caption{ Convergence rate of ASCI and ASCI+PT2 energies for the CN radical, with increasing size of ASCI wave functions. The reference energy $E_{ref}$ is the ASCI+PT2 energy with 1 million determinants in the variational wave function, along with orbital rotations. We observe that ASCI+PT2 energies with as few as 1000 determinants are more accurate than the "gold standard" CCSD(T) results. The PT2 contribution is essential for this enhanced accuracy, and should be employed whenever possible. Natural orbital rotations are also found to reduce the error to a great extent, and are recommended at all times.}
\label{fig:cn-ene}
\end{figure}

We also make comparison of the ASCI results with CCSD(T)\cite{raghavachari1989fifth} and CCSDTQ. Before discussing the full benchmark set, we demonstrate in Figure~\ref{fig:cn-ene} the convergence of ASCI for the cyanide radical (CN) and compare this with convergence of a set of coupled cluster simulations. The coupled cluster results, for both CCSD and CCSD(T), were performed in QChem 5.1~\cite{shao_advances_2015}. As seen in this figure, the ASCI + PT2 results are more accurate than the comparable CCSD(T) results, even for calculations with only $10^{3}$ determinants for the variational wave function. Thus while simulations of fully converged ASCI calculations can take a few hours, small simulations that only take a few minutes (or less) already show similar or better accuracy than CCSD(T).

In the rest of this section we discuss the  energies 
from our simulations that can serve as benchmarks for future methodological developments. Tables~\ref{tab:dz} (cc-pVDZ) and Table~\ref{tab:tz} (cc-pVTZ) present the variational energies and the perturbation results from ASCI, as well as comparisons to CCSD(T). Table~\ref{tab:atom} presents  atomization energies and makes comparisons of these to values obtained from CCSD(T) and CCSDTQ. 

 The results for the cc-pVDZ basis set in Table~\ref{tab:dz} show apparent convergences of the energy  to below 1 mHa accuracy for all molecules. It is possible to extrapolate such results when extra accuracy is needed~\cite{holmes2016}. We note that when the perturbation correction is added to ASCI, neither this nor CCSD(T) is guaranteed to be variational. However, with the exception of Si$_2$H$_6$, all ASCI PT2  energies are below the CCSD(T) energies.
   We suggest that the difference between ASCI and CCSD(T) is a good estimate for the error in CCSD(T) energies relative to the full configuration interaction (FCI) result. In general these differences are less than 1 mHa. However Table~\ref{tab:tz} shows that for several molecules it can be higher, with the largest difference being for the CN radical, which has a difference of 4.2 mHa. For many of the molecules, either the Hilbert space is small or the difference between the variational and PT2 estimates is less than 1 mHa for small N$_{tdets}$, and in these cases we do not require testing of convergence with up to 1 million determinants.

 The results for the cc-pVTZ basis set (Table~\ref{tab:tz}) indicate that the ground state energies of many of the molecules are also converged in a larger basis set. 
For molecules not already converged with $N_{dets}$ = 100,000, we were able to run PT2 corrections with 100,000 and 300,000 determinants. The largest calculation made was for Si$_2$H$_6$ with 300,000 determinants, for which extra computing time was needed to converge the simulation. We note that in contrast to results from ASCI in comparing to the cc-pVDZ basis set energies,
several CCSD(T) energies lie below the ASCI results. The full significance of this will have to be investigated further with larger determinant calculations in the future. For many systems, the CCSD(T) error is very similar across the basis sets and does not actually get worse when comparing cc-pVDZ to cc-pVTZ.
 
\begin{figure*}
\begin{center}
\scalebox{1}{\includegraphics[width=2.0\columnwidth]{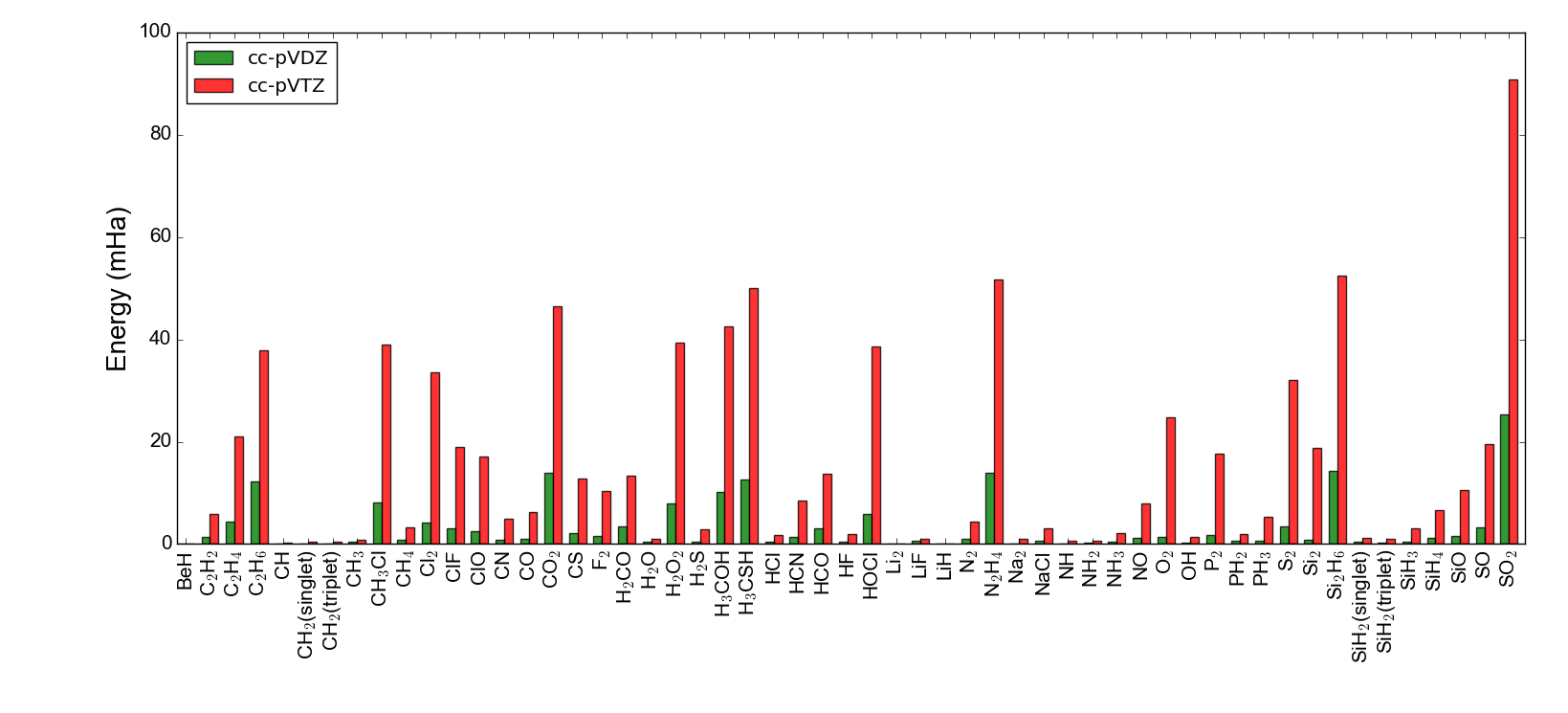}}
\end{center}
\caption{Plot showing the PT2 energies for the G1 set, calculated with cc-pVDZ and cc-pVTZ basis sets. The PT2 energy is obtained from the most accurate calculation performed for each system. For specific values and details, see Table~\ref{tab:dz} for the cc-pVDZ simulations and Table~\ref{tab:tz} for the cc-pVTZ simulations. }
\label{fig:pt2}
\end{figure*}

 Another way to understand the convergence errors in ASCI is to look at the size of the PT2 energy corrections. These are shown for the G1 dataset in figure \ref{fig:pt2}.  Within the G1 set, the SO$_{2}$ molecule has the largest PT2 correction.  We note that while Si$_2$H$_6$ is one of the most time-consuming simulations here, it does not have the largest PT2 correction.  
 
 Table~\ref{tab:atomene} presents atomic ground state energies for the atoms contained in the G1 set that are obtained from ASCI with cc-pVDZ and cc-pVTZ basis sets. These are combined with the molecular data in Tables~\ref{tab:dz} and~\ref{tab:tz} to calculate the atomization energies for the molecules in the G1 set, which are shown in Table~\ref{tab:atom} and compared there to other benchmark simulations. For the cc-pVDZ results (columns 2 and 7), we compare with results from CCSDTQ~\cite{feller1}. These comparisons are necessarily indirect, since different geometries were used between the two sets of results, and the ASCI results are all electron. However the difference in geometries are not so large.  Surveying the full set of molecules, we find less than 1 kcal differences between atomization energies.  For the cc-pVTZ results (columns 4 and 9), we compare to benchmark complete basis set (CBS) frozen core results~\cite{feller1,feller2}. Again the comparison is not direct, but it is clear that the atomization energies are trending in the right direction in comparison to the CBS limit.  To demonstrate the similarity of geometries between the different calculations, in Table~\ref{tab:feller} we present results of ASCI with perturbation corrections for selected molecules using the geometries from Feller \textit{et. al.}~\cite{feller1}.  These can be compared with the corresponding energies from Table~\ref{tab:dz} and \ref{tab:tz} to estimate the energy difference between the two sets of geometries and confirm the similarity.  

It is also interesting to compare with recent auxiliary field quantum Monte Carlo results (AFQMC) that are now able to calculate the G1 set~\cite{borda2018}. %This is a significant advancement, as some of the cc-pVTZ simulations presented here have more than 30 electrons and more than 150 orbitals, which is beyond what can routinely be calculated with both DMRG and FCIQMC for chemical systems.  
AFQMC is accurate on many of these systems, but 
loses accuracy in certain situations such as for ionic systems like LiF and NaCl.  This is likely due to deficiencies in the current trial wave functions used in AFQMC.  See Ref.~\cite{borda2018} for more details. 
Further comparisons between these methods will be interesting in the future. We also note that recent work on DMRG has also added perturbative corrections to that methodology~\cite{guo2018-1,guo2018-2}, although it is not clear whether the largest simulations required for
the G1 set are feasible with DMRG, even with perturbation theory improvements.  Entanglement in chemical systems can be quite large even in systems in which the many body interactions are not necessarily strong, and is thus a problem for efficient simulation with DMRG.

\section{Conclusions}
%\textit{Conclusions}:
In this work we have presented a number of new algorithms and techniques for performing selected CI simulations and integrated them into an updated ASCI algorithm.  These techniques include new ASCI search, dynamic bit masking, fast orbital rotations, and fast diagonal matrix elements. We also presented other techniques, such as residue arrays and alternative bit string representations, which will be important for simulating certain types of Hamiltonians. 
%These algorithms are not limited to selected CI, and are likely to be useful in other related CI algorithms. 

Most of the new techniques presented here take advantage of modern sorting techniques that have been designed to be efficient on current computing architectures.
Sorting based algorithms provide new avenues for using GPUs and parallel sorting on CPUs.  Sorting and hashing techniques can be further combined together, to allow calculations with very large basis sets.

Our results indicate that the resulting algorithms are faster than other previously published selected CI approaches. 
To make comparison with stochastic approaches to CI such as HBCI,
we have presented initial benchmarking calculations for C$_{2}$ and F$_{2}$ in Table~\ref{tab:hbcomp} that show the new ASCI approach is about an order of magnitude faster 
than HBCI
for target accuracies within 0.1 mHa  and more than two orders of magnitude faster for target accuracies within 0.01 mHa.

With the ideas presented in this work, we have demonstrated the importance of using modern algorithms and  understanding computing architectures for design of efficient selected CI algorithms.  
The speed and accuracy of these new techniques over a wide range of systems were demonstrated with a full set of ground state calculations for the G1 dataset of small molecules. We presented comparisons to coupled cluster (CCSD(T) and CCSDTQ) and benchmarked the accuracy of this compared to ASCI. This demonstration shows the various convergence properties and the accuracy that can be expected from the latest ASCI approach today.  Given the results presented here, it is evident that we are now at a new beginning for modern selected CI simulations and that many hitherto intractable new applications are now within reach.

\section{Acknowledgements}
This work was supported through the Scientific Discovery through
Advanced Computing (SciDAC) program funded by the U.S. Department of
Energy, Office of Science, Advanced Scientific Computing Research and
Basic Energy Sciences.  We used the Extreme Science and Engineering Discovery
Environment (XSEDE), which is supported by the National Science Foundation Grant No. OCI-1053575. CDF was supported by the NSF Graduate Research Fellowship under Grant DGE-1106400 and by the DOE Office of Science Graduate Student Research (SCGSR) program under contract number DESC0014664. DH was supported by a Berkeley Fellowship. DL and MHG acknowledge support from the Director, Office of Science, Office of Basic Energy Sciences, of the U.S. Department of Energy under Contract No. DE-AC02-05CH11231.

\section{Supplemental Information for the Chromium Dimer}
\begin{figure}
\begin{center}
\scalebox{1}{\includegraphics[width=1.0\columnwidth]{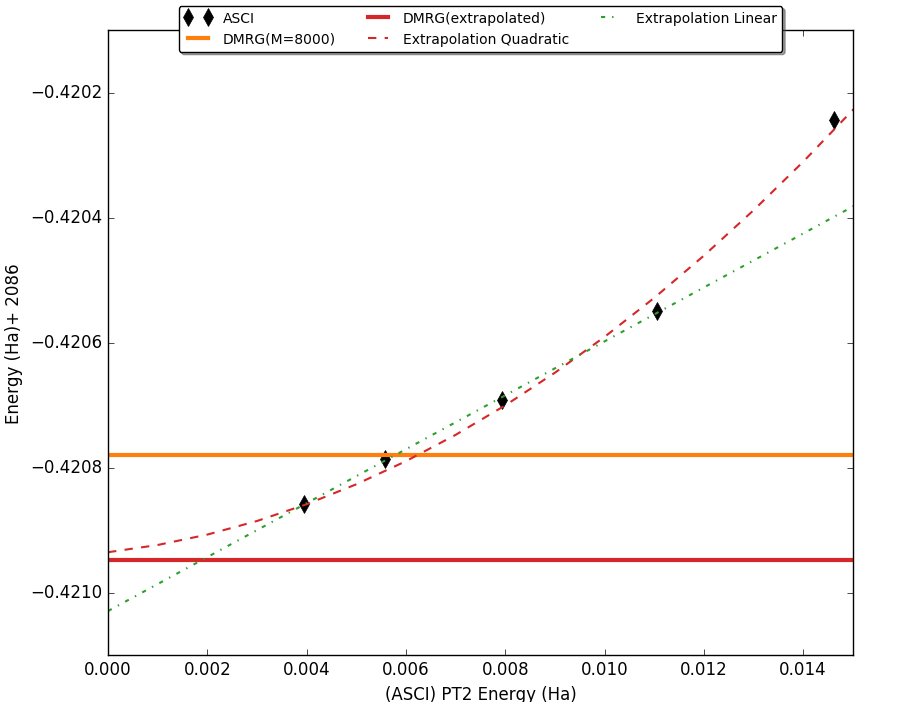}}
\end{center}
\caption{\textcolor{black}{A plot of Cr$_{2}$ SVP (24e,30o) energies using ASCI search to generate highly converged wave functions.  The data points correspond to wave functions with 0.3, 1.0, 2.0, 4.0 and 8.0 million determinants.   We compare with the best published DMRG values from ref.~\cite{amaya2015}, and our ASCI+PT2 values for Cr$_{2}$.  Also plotted is the extrapolated DMRG result (-2086.420948 Ha).  For the ASCI data  we plot a linear extrapolation to the best four ASCI values and in addition we also plot a quadratic extrapolation to all our data.  The linear extrapolation  yields (-2086.421029 Ha), and the quadratic extrapolation yields (-2086.420935 Ha).}}
\label{fig:cr2ext}
\end{figure}

In this section we present an extrapolation for the ground state energy of the Cr$_{2}$ in the SVP basis (24e,30o) and make a comparison to previously published DMRG results.  The energies and extrapolations are presented in Figure~\ref{fig:cr2ext} and show excellent agreement between the methods.

\newpage
\appendix
\section{Hartree-Fock energies for G1 set molecules}
The energies of the restricted Hartree-Fock determinants used to initialize the ASCI calculations are given in Tables \ref{tab:hf-all} and \ref{tab:hf-feller}. Stability analysis was performed to ensure that all SCF solutions were local minimas.  
\begin{table}[htb!]
\centering
\footnotesize
\begin{tabular}{|l|l|l|}
\hline
System       & cc-pVDZ      & cc-pVTZ      \\ \hline
BeH           & -15.14945189 & -15.15205222 \\ \hline
C$_2$H$_2$          & -76.82472747 & -76.84763525 \\ \hline
C$_2$H$_4$          & -78.03990264 & -78.06354855 \\ \hline
C$_2$H$_{6}$          & -79.23494277 & -79.25973681 \\ \hline
CH            & -38.26877842 & -38.27690073 \\ \hline
CH$_2$\_singlet  & -38.88108554 & -38.89231068 \\ \hline
CH$_2$\_triplet  & -38.92139886 & -38.93217379  \\ \hline
CH$_3$           & -39.55963482 & -39.57296784 \\ \hline
CH$_3$Cl         & -499.1177293 & -499.1476064 \\ \hline
CH$_4$           & -40.19870854 & -40.21331465 \\ \hline
CN            & -92.19603079 & -92.21741416  \\ \hline
CO            & -112.7461016 & -112.7766305 \\ \hline
CO$_2$           & -187.6463113 & -187.7018166 \\ \hline
CS            & -435.3293719 & -435.3521612 \\ \hline
Cl$_2$           & -918.9609562 & -918.9983715 \\ \hline
ClF           & -558.8436442 & -558.8989343 \\ \hline
ClO           & -534.2516809 & -534.2964982 \\ \hline
F$_2$            & -198.6847963 & -198.7508413 \\ \hline
H$_2$CO          & -113.8746242 & -113.9098452 \\ \hline
H$_2$O           & -76.02602772 & -76.05613647 \\ \hline
H$_2$O$_2$          & -150.7817569 & -150.8331176 \\ \hline
H$_2$S           & -398.6946587 & -398.7129979 \\ \hline
H$_3$COH         & -115.0486003 & -115.0884306 \\ \hline
H$_3$CSH         & -437.7255248 & -437.756402  \\ \hline
HCN           & -92.87969951 & -92.90351433 \\ \hline
HCO           & -113.2518818 & -113.2848402 \\ \hline
HCl           & -460.0894453 & -460.1067487 \\ \hline
HF            & -100.0184682 & -100.0569205 \\ \hline
HOCl          & -534.8720321 & -534.9177714 \\ \hline
Li$_{2}$           & -14.87000212 & -14.87168961 \\ \hline
LiF           & -106.9451499 & -106.9801182 \\ \hline
LiH           & -7.98363507  & -7.986532056 \\ \hline
N$_2$            & -108.9466732 & -108.9743976 \\ \hline
N$_2$H$_4$          & -111.1858923 & -111.2234062 \\ \hline
NH            & -54.95953403 & -54.97346363 \\ \hline
NH$_2$           & -55.56273484 & -55.58092764 \\ \hline
NH$_3$           & -56.19548576 & -56.21749393 \\ \hline
NO            & -129.254714  & -129.2901963 \\ \hline
Na$_{2}$           & -323.7047434 & -323.7151846 \\ \hline
NaCl          & -621.4337887 & -621.4537044 \\ \hline
O$_2$            & -149.6014308 & -149.6460477 \\ \hline
OH            & -75.3896954  & -75.41403106 \\ \hline
P$_2$            & -681.4629985 & -681.4863607 \\ \hline
PH$_2$           & -341.8675755 & -341.8815522 \\ \hline
PH$_3$           & -342.4706082 & -342.4876397 \\ \hline
S$_2$            & -795.0490094 & -795.0805171 \\ \hline
SO            & -472.330077  & -472.3804324 \\ \hline
SO$_2$           & -547.1725083 & -547.2750501 \\ \hline
Si$_2$           & -577.756499  & -577.7739636 \\ \hline
Si$_2$H$_{6}$         & -581.338866  & -581.3706463 \\ \hline
SiH$_2$\_singlet & -290.0184384 & -290.0303487 \\ \hline
SiH$_2$\_triplet & -289.9952407 & -290.0072341 \\ \hline
SiH$_3$          & -290.6235912 & -290.6386838 \\ \hline
SiH$_4$          & -291.2428929 & -291.2605406 \\ \hline
SiO           & -363.7883502 & -363.8356554 \\ \hline
\end{tabular}
\caption{Restricted Hartree-Fock energies for molecules in the G1 set (RHF for closed shell species and ROHF for open shell).  Energies are presented in units of Ha. The geometries are the same as those employed in Tables \ref{tab:dz} and \ref{tab:tz} (i.e. are taken from the original G1 set~\cite{pople1989} except for CN and CH$_{2}$ triplet, for which we use the geometries from Ref.~\cite{feller1}.)}
\label{tab:hf-all}
\end{table}

\begin{table}[htb!]
\centering
\footnotesize
\begin{tabular}{|l|l|l|}
\hline
System       & cc-pVDZ      & cc-pVTZ      \\ \hline
C$_2$H$_4$  & -78.03884396 & -78.0628233  \\ \hline
C$_2$H$_6$  & -79.23488258 & -79.25993592 \\ \hline
CH$_3$Cl & -499.1177629 & -499.1478118 \\ \hline
CH$_4$   & -40.19865157 & -40.21342364 \\ \hline
Cl$_2$   & -918.9608245 & -918.9987691 \\ \hline
ClF   & -558.8442574 & -558.9009543 \\ \hline
ClO   & -534.2495837 & -534.295923  \\ \hline
CO    & -112.7480967 & -112.7789016 \\ \hline
CO$_2$   & -187.6511077 & -187.7072565 \\ \hline
CS    & -435.3296827 & -435.3526846 \\ \hline
F$_{2}$    & -198.6856445 & -198.7520074 \\ \hline
H$_2$CO  & -113.8763719 & -113.9119201 \\ \hline
H$_2$O$_2$  & -150.78415   & -150.8360866 \\ \hline
H$_2$S   & -398.694525  & -398.7129408 \\ \hline
HCN   & -92.88324631 & -92.90805477 \\ \hline
HCO   & -113.2530034 & -113.28603   \\ \hline
N$_2$    & -108.9541417 & -108.9834898 \\ \hline
NH$_3$   & -56.19563482 & -56.21791293 \\ \hline
NO    & -129.2536412 & -129.2889892 \\ \hline
P$_2$    & -681.465778  & -681.4896859 \\ \hline
PH$_3$   & -342.4703141 & -342.4873996 \\ \hline
Si$_2$   & -577.7566459 & -577.7741915 \\ \hline
SiO   & -363.7899104 & -363.8389006 \\ \hline
SO    & -472.3332171 & -472.386195  \\ \hline
SO$_2$   & -547.1791953 & -547.2876407 \\ \hline
\end{tabular}
\caption{Restricted Hartree-Fock energies (RHF for closed shell species and ROHF for open shell) for selected molecules using geometries from Feller \textit{et. al.}~\cite{feller1}.}
\label{tab:hf-feller}
\end{table}

\begin{figure*}
\centering	\includegraphics[width=10cm, height=4.5cm]{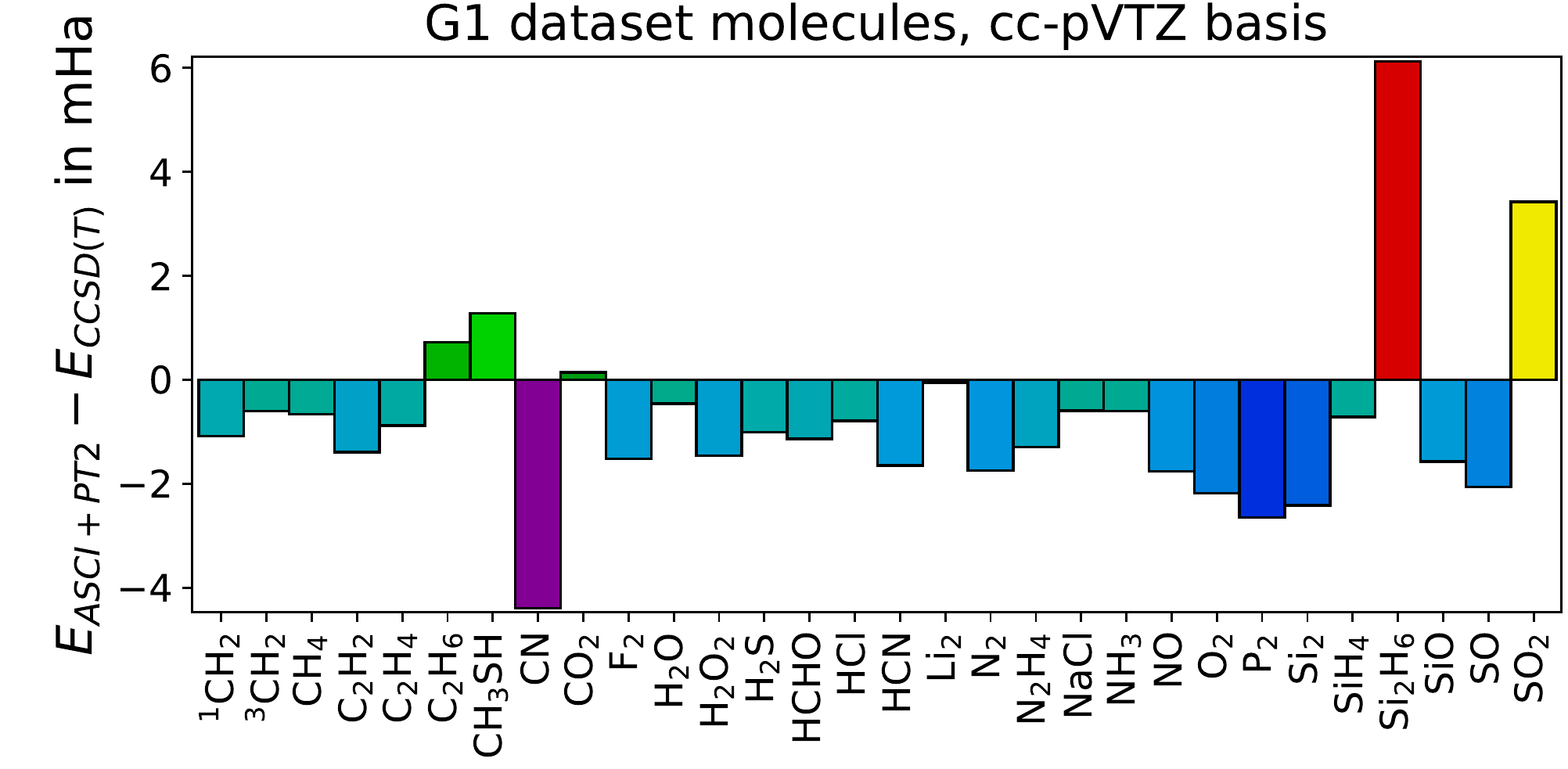}
\caption{Table of Contents Figure}
\end{figure*}

%%%%%%%%%%%%%%%%%%%%%%%%%%%%%%%%%%%%%%%%%%%%%%%%%%

%\bibliography{refs}
%\bibliography{refs}

%

\end{document}